\documentclass[letterpaper]{JHEP3}
\pdfoutput=1
\usepackage{amsmath}
\usepackage{epsfig}

\DeclareMathOperator{\e}{e} \DeclareMathOperator{\const}{const}
\DeclareMathOperator{\sgn}{sgn}

\newcommand{\be}{\begin{equation}}
\newcommand{\ee}{\end{equation}}
\newcommand{\beq}{\begin{equation}}
\newcommand{\eeq}{\end{equation}}
\newcommand{\beqa}{\begin{align}}
\newcommand{\eeqa}{\end{align}}

\newcommand{\dbar}{{\mathchar'26\mkern-11mu\mathrm{d}}}

\renewcommand{\geq}{\geqslant}

\newcommand{\xb}{\boldsymbol{x}}

\newcommand{\ds}{\mathrm{d}}

\newcommand{\ms}{\mathrm{m}}

\newcommand{\ssl}{\mathrm{s}}

\newcommand{\Lcal}{\ensuremath{\mathcal{L}}}

\newcommand{\Ncal}{\ensuremath{\mathcal{N}}}

\newcommand{\Pcal}{\ensuremath{\mathcal{P}}}

\newcommand{\Ecal}{\ensuremath{\mathcal{E}}}

\newcommand{\intfb}[1]{\ensuremath{\int\dbar^4k \,}}
\newcommand{\inttb}[1]{\ensuremath{\int\dbar^3k \,}}

\newcommand{\Mpl}{M_\text{Pl}}

\newcommand{\proj}{\perp}

\newcommand{\lp}{\left}
\newcommand{\rp}{\right}

\newcommand{\ext}{\text{ext}}

\title{Imperfect Dark Energy from Kinetic Gravity Braiding}
\author{C\'edric Deffayet\\AstroParticule \& Cosmologie, UMR7164-CNRS,
Universit\'e Denis Diderot-Paris 7, CEA, Observatoire de Paris,
10 rue Alice Domon et L\'eonie Duquet, \\ F-75205 Paris Cedex 13, France\\ \email{deffayet@iap.fr}}

\author{Oriol Pujol\`as\\CERN, Theory Division, CH-1211 Geneva 23, Switzerland \\ \email{oriol.pujolas@cern.ch}}

\author{Ignacy Sawicki$^a$ and Alexander Vikman$^b$\\Center for Cosmology and Particle Physics, New York University, \\ New York, NY 10003, USA \\
$^a$\email{ignacy.sawicki@nyu.edu} \\ $^b$\email{alexander.vikman@nyu.edu}}
\preprint{CERN-PH-TH/2010-166}
\keywords{dark energy theory; modified gravity; scalar-tensor models; galileon}
\abstract{We introduce a large class of scalar-tensor models with interactions containing the second derivatives of the scalar field but not leading to additional degrees of freedom. These models exhibit peculiar features, such as an essential mixing of scalar and tensor kinetic terms, which we have named \emph{kinetic braiding}. This braiding causes the scalar stress tensor to deviate from the perfect-fluid form. Cosmology in these models possesses a rich phenomenology, even in the limit where the scalar is an exact Goldstone boson. Generically, there are attractor solutions where the scalar monitors the behaviour of external matter. Because of the \emph{kinetic braiding}, the position of the attractor depends both on the form of the Lagrangian and on the external energy density. The late-time asymptotic of these cosmologies is a de~Sitter state. The scalar can exhibit phantom behaviour and is able to cross the phantom divide with neither ghosts nor gradient instabilities. These features provide a new class of models for Dark Energy. As an example, we study in detail a simple one-parameter model. The possible observational signatures of this model include a sizeable Early Dark Energy 
and a specific equation of state evolving into the final de-Sitter state from a healthy phantom regime.}

\begin{document}
\section{Introduction, Summary and Future Directions}

Infrared modifications of gravity have received considerable attention recently for the potentially natural explanation of the current cosmic acceleration which they might offer. Much advance has been accomplished, with several consistent models now well understood, such as (the normal branch of) DGP-type models \cite{Dvali:2000hr,deRham:2007xp} as well as a number of Lorentz-violating setups 
\cite{ArkaniHamed:2003uy,Rubakov:2004eb,Dubovsky:2004sg}, see also review \cite{Rubakov:2008nh}. An important and intimately related spinoff from these investigations has been the study of 
consistent higher derivative interactions for scalar fields \cite{lpr,Nicolis:2008in,covGal}.
In many modified gravity models the graviton is massive, which implies the presence of additional polarizations, the most important of which is a scalar longitudinal mode. This polarization can be viewed as the `Goldstone of a Goldstone' \cite{ArkaniHamed:2002sp}, and as a consequence the action for this mode tends to involve its second derivative. This is at the heart of the inherent difficulty with massive gravity models, since second-derivative interactions easily lead to a higher derivative equation of motion and, hence, to ghosts at nonlinear level \cite{ArkaniHamed:2002sp,Deffayet:2005ys,Creminelli:2005qk}. By exploring the decoupling limit of DGP, in \cite{lpr} (see also \cite{nr}) it was realized that the leading cubic interaction for the scalar $\phi$ is of the form $(\partial\phi)^2\Box\phi$. This interaction leads to an equation of motion containing exclusively derivatives of second order, and hence to a ghost-free implementation of the Vainshtein effect \cite{lpr,nr,Deffayet:2005ys,Babichev:2009ee, Babichev:2009us, Babichev:2010jd}. More recently, two more interaction terms for a scalar field (quadratic and cubic in $\partial\partial\phi$) leading to equations of motion of second order have been discussed\footnote{Similar models were introduced before in another context \cite{Fairlie:1991qe,Fairlie:1992nb}.  Also, recently, such models have been extended to vectors and $p$-forms \cite{Deffayet:2010zh}.} \cite{Nicolis:2008in}, along with their covariantization \cite{covGal,Deffayet:2009mn}. See also Refs.~\cite{deRham:2010eu, deRham:2010gu, deRham:2010ik} where the investigation of a consistent theory of massive gravity in the decoupling limit eventually leads to the same additional terms in the Lagrangian. 

Our purpose in this article is to abstract these developments in modified gravity and to consider the addition of second-derivative couplings as a generalisation of scalar-tensor theories. Specifically, we explore a rather large family of theories where the Lagrangian for the scalar field takes the form
\be\label{L}
{\cal L}=K(\phi,X)+G(\phi,X)~\Box\phi~,
\ee
with $X\equiv \frac{1}{2}g^{\mu\nu}\partial_\mu\phi\partial_\nu\phi$ and $K,G$ some generic functions, with $K$ proposed already as k-\emph{essence} \cite{ArmendarizPicon:1999rj, ArmendarizPicon:2000dh, ArmendarizPicon:2000ah}. We will show that for any form of the functions  $K,G$ the equations of motion are of second order.\footnote{Generic choices of $K,G$ give equations of motion with explicit dependence on the first derivative $\nabla_\mu\phi$. Hence, the Lagrangians \eqref{L} do not have the Galilean symmetry which characterizes the interactions of \cite{lpr} and \cite{Nicolis:2008in}.} 
 
Hence, this Lagrangian can encode perfectly valid interactions for a single scalar degree of freedom. 
As we will see, a theory like \eqref{L} has rather unfamiliar properties which we will elucidate. We will concentrate on applying such constructions to model Dark Energy. 

The Lagrangian \eqref{L} can be viewed as reminiscent of a partial resummation of a gradient expansion commonly used in fluid dynamics. As is well known, the stress tensor for k-\emph{essence} (the $G=0$ case) takes the perfect-fluid form. As we will elaborate on in a future work \cite{Imperfect}, the $G\Box\phi$ interactions give rise to deviations of a certain type from the perfect-fluid picture. Hence, we will refer to the scalar field $\phi$ described by \eqref{L} as an \emph{imperfect} scalar.

Let us also remark on another interesting property of the models of type \eqref{L}. Once gravity is introduced, the interaction term $G \Box\phi$ contains a scalar-tensor kinetic coupling, schematically $G\,\partial g \partial\phi$, due to the Christoffel symbols present in the covariant D'Alembertian operator. A consequence of these couplings is a novel kind of mixing which we have named {\em kinetic braiding}. It implies that the scalar equation of motion contains the second derivative of the metric---and vice-versa---in an essential way: there exists no Einstein frame where the kinetic terms are diagonalised\footnote{An ordinary scalar with nonminimal coupling to curvature, e.g.\ $\phi^2R$, also leads to kinetic mixing. However, in this case the mixing is non-essential in the sense that it can be removed by a field redefinition (in the form of a conformal transformation). Note that \emph{kinetic braiding} is present in the higher-order terms of the covariant Galileon \cite{covGal,Deffayet:2009mn}.}. This \emph{kinetic braiding} manifests itself either at nonlinear level or on non-trivial backgrounds. In this sense, this leads to a novel modification of gravity that perhaps has not been fully appreciated.  Another consequence is the fact that this operator may be non-zero and actually important on backgrounds where the second derivatives vanish, e.g.\ the sound speed can be non-vanishing in an exact de-Sitter spacetime. We will give some examples of these in the text. 

An obvious domain in which models of type \eqref{L} can have interesting applications is cosmology.  As a consequence of \emph{kinetic braiding}, the Friedmann equation has remarkable properties, such as the presence of a term linear in $H$, similarly to the DGP model. Moreover, even in the absence of any direct coupling between the $\phi$ and external matter in the action, the evolution of the scalar $\phi(t)$ generically \emph{monitors} any form of external energy density $\rho_{\ext}$ present. In the shift-symmetric case, where $\phi$ is a Goldstone boson, there exist attractor solutions on which the shift charge exactly vanishes such that the energy density stored in the scalar is a certain model-dependent function of the external energy density, $\rho_{\ext}$: $\rho_\phi=\rho_\phi^*(\rho_{\ext})$. 
Furthermore, for appropriately chosen $K,G$ (whenever $\rho_\phi^*(0)>0$), the attractor tends to de Sitter space in the asymptotic future. 
This is similar to ghost condensation \cite{ArkaniHamed:2003uy}, 
the difference being that on the attractor the ghost condensate behaves like a cosmological constant. On the other hand, the \emph{imperfect} scalar has a nontrivial equation of state dependent on the external matter content. 

The existence of this monitoring behaviour opens up a number of applications, the most obvious of which is that $\phi$ can play the role of dark energy. Under very general circumstances, the on its attractor $\phi$ behaves as a phantom, $w_X^*<-1$. We will show that when evolution is started away from the attractor, the imperfect scalar initially redshifts in the usual manner. Once the attractor is approached, $\phi$ crosses the phantom divide and ends up dominating and accelerating the universe at late times. All of this can be accomplished without generating ghosts or gradient instabilities.  

The existence of a physically plausible theory that can cross the phantom divide, i.e.\ violate the Null Energy Condition, opens a Pandora's box full of exotic phenomena such as wormholes, bouncing universes, etc. However, currently, $w_{X}<-1$ is not observationally excluded, see e.g.~\cite{Komatsu:2010fb}. 

We analyse in detail a simple one-parameter example that exhibits the features above, where the \emph{imperfect} scalar on the attractor evolves from a subdominant phantom (with $w_X=-2$) during matter domination to a dominating late-time de Sitter phase. In addition, the energy density of this fluid while off-attractor, has a local maximum at matter-radiation equality, allowing us to propose this model as a simple example of a form of early dark energy.

This article is organized as follows: We introduce the class of models \eqref{L} in section~\S\ref{s:genprop}, discussing their equations of motion and coupling to gravity along with the effective metric for the propagation of perturbations. We specialise our results to the cosmological background in section~\S\ref{s:cosmo}, and discuss the stability conditions for background solutions: positivity of the sound speed and of the kinetic term of perturbations. In section~\S\ref{s:shift}, we specialise to the cosmology of the shift-symmetric case. We discuss the dynamical attractors in the presence of matter and in the late universe---once matter has diluted away. In section~\S\ref{s:linmodel}, we discuss a simple one-parameter model to demonstrate explicitly the sort of phenomenology one can obtain within the new framework proposed. We invite those readers more interested in dark-energy phenomenology to start there. Finally, in Appedix~\ref{sec:apppert}, we present a derivation of the quadratic action for cosmological perturbations for theories described by \eqref{L}.

This work has a certain overlap with Refs~\cite{Chow:2009fm, Silva:2009km, Kobayashi:2010wa, DeFelice:2010jn, DeFelice:2010gb} where the cosmology of the decoupling limit of DGP and some of its simple modifications are considered.  These references concentrate on models with $G=Xf(\phi)$ and non-minimally coupled to the Ricci scalar, \`{a} la Brans-Dicke gravity. In this work, we extend the above class of theories to a general function $G(\phi,X)$ while restricting to a minimal coupling to gravity. As such we make it clear that the new interesting phenomenology is a result of \emph{kinetic braiding}. In \cite{Creminelli:2010ba,Nicolis:2009qm} the authors take the effective field theory approach to discuss the ability of generalisation of the decoupling limit of DGP to stably violate the Null Energy Condition and propose a new mechanism to replace inflation as the initial stage of the universe. In addition, in Refs~\cite{DeFelice:2010pv, Gannouji:2010au}, the cosmology of higher-order terms of the general covariant galileon is analysed. 

We believe that the models described by Eq.~\eqref{L} provide many avenues for exploration. In particular, the reinterpretation of these models as imperfect fluids yields many interesting insights, as we will show in our upcoming work, Ref.~\cite{Imperfect}. It is well known that the classical results of single-field inflation can be obtained entirely in terms of perfect-fluid variables \cite{Garriga:1999vw}. The implementation of inflation using \emph{imperfect} scalars provides another rich arena for model-building, with potentially non-trivial effects on cosmological perturbations. We will return to a thorough investigation of cosmological perturbations in the future.


\section{General Properties of Kinetic Gravity Braiding Theories}
\label{s:genprop}
In this paper we consider a class of scalar field theories minimally
coupled to gravity, which is described by the action
\footnote{We use the metric signature convention $\left(+-\,-\,-\right)$.
} 
\begin{equation}
S=S_{\text{EH}}+S_{\phi}=\int\mbox{d}^{4}x\sqrt{-g}\left[-\frac{R}{2}+K+G\times B\right],\label{e:action}
\end{equation}
where we denote 
\begin{eqnarray*}
X\equiv\frac{1}{2}g^{\mu\nu}\nabla_{\mu}\phi\nabla_{\nu}\phi, & \text{and} & B\equiv g^{\mu\nu}\nabla_{\mu}\nabla_{\nu}\phi\equiv\Box\phi\,,
\end{eqnarray*}
and $K\left(\phi,X\right)$ and $G\left(\phi,X\right)$ are arbitrary
functions of the scalar field and its standard kinetic term $X$.
Here and in most of the paper, we use reduced Planck units where $M_{\text{Pl}}=\left(8\pi G_{\text{N}}\right)^{-1/2}=1$.
In further discussion we will also use 
the corresponding Lagrangian for the scalar \[
\mathcal{L}\left(\phi,X,B\right)=K\left(\phi,X\right)+G\left(\phi,X\right)\times B\,,\]
which we will consider as a function of three variables: $\phi$, $X$ and
$B$; whenever we differentiate with respect to $\phi$ or $X$, we are keeping $B$ constant. Our class of models extends those studied up to now by generalising the interaction term $X\box\phi$ of DGP decoupling limit and the galileon models to a function $G(\phi,X)\box\phi$, significantly enriching the phenomenology. 
  
One can naively motivate our interest in non-canonical field theories
containing the new $G\times B$ term by a similar argument to that
for k-\emph{essence}. In k-\emph{essence}, we treat the canonical kinetic term $X$
as a first term in an effective-field-theory expansion, since no symmetries
exist that would prevent additional contributions, i.e.\ \[
K\left(\phi,X\right)\sim X\left(1+c_{1}\left(\phi\right)X+c_{2}\left(\phi\right)X^{2}+...\right)\,.\]
Then after resummation of these corrections one obtains a non-canonical
Lagrangian with a derivatively self-coupled scalar field, e.g.\ as in the case
of the Dirac-Born-Infeld theories. However, one
could equally well start from the canonical kinetic term $X$ and integrate it
by parts with no effect on the observed dynamics. In that case, it
would not be unreasonable to expect there to exist higher-order
corrections similar to those discussed above, \[
G\left(\phi,X\right)\Box\phi\sim-\phi\Box\phi\left(1+\tilde{c}_{1}\left(\phi\right)X+\tilde{c}_{2}\left(\phi\right)X^{2}+...\right)\,.\]
Note that both these expansions are in $X$. Below
we will show that it is not possible to recast such corrections to
$-\phi\Box\phi$ as a modification of the function $K\left(\phi,X\right)$
and surface terms in the action, and therefore these theories are
outside of the k-\emph{essence} scheme.  We stress that, in the Einstein-Hilbert action, the second derivative enters
the Lagrangian only linearly and, despite
the presence of these higher derivatives in the Lagrangian, the equations
of motion remain of the second order. Therefore, similarly to the
Einstein's general relativity, there are no new (ghosty) degrees
of freedom which would appear as the result of the Ostrogradsky theorem\footnote{There is another important similarity with General Relativity: a rigorous formulation of the variational principle requires a boundary term similar to the Gibbons-Hawking-York term in General Relativity. In this work we do not consider this boundary term assuming that it exists. The term appropriate for the decoupling limit of DGP was found in \cite{Dyer:2009yg}.}. Thus, at least for some functions $K\left(\phi,X\right)$ and $G\left(\phi,X\right)$
and some general backgrounds, there are no ghosts and pathologies associated with
them---the theories containing $G\times B$ are
on the same footing as k-\emph{essence} or DBI theories. 

The form of the action in Eq.$\,$\eqref{e:action} can be further
generalised by introducing a non-minimal coupling of the scalar field
to gravity, or a modification of the $f(R)$ type. An appropriate
change of variables will transform the action to the Einstein frame, where the
usual Einstein-Hilbert term is restored. In the case of a theory containing
only gravity and the scalar, this is equivalent to choosing different
functions $K$ and $G$. However, if external matter is present, this
transformation introduces an additional coupling between the matter
and the scalar field\footnote{Following the same logic one could also consider additional derivative
couplings of the scalar field to the Einstein tensor. In that case, equations
of motion would still be of the second order, see e.g.\ recent works
\cite{Sushkov:2009hk, Saridakis:2010mf, Gao:2010vr, Germani:2010gm, Germani:2010ux}, contrary to \cite{Mukohyama:2003nw, Nojiri:2004bi}.}.  Such theories can actually evade Solar-System tests of general relativity
because of the presence of the Vainshtein mechanism \cite{Babichev:2009ee,Babichev:2009jt}. However, in this work we will concentrate on the minimally coupled models. This minimal coupling
to gravity is the crucial difference of our work from the recent studies
of the Galileon scalar-tensor models \cite{Chow:2009fm,Silva:2009km,Kobayashi:2010wa,DeFelice:2010jn,DeFelice:2010gb,Kobayashi:2009wr}.
In particular, we have to stress that the most unexpected and interesting
features of these theories are a result not only of the possible non-minimal
coupling of the scalar to gravity but of the presence of $G\times B$
in the action. 

In our discussion we will frequently use a Lagrangian
equivalent to $\mathcal{L}$: integrating by parts the scalar-field
contribution $S_{\phi}$ to the action Eq.~\eqref{e:action}, we
obtain 
\begin{equation}
\mathcal{P}=K-\left[\left(\nabla^{\lambda}\phi\right)\nabla_{\lambda}\right]G=K-2XG_{\phi}-G_{X}\nabla^{\lambda}\phi\nabla_{\lambda}X\,,\label{e:L2}
\end{equation}
where the subscripts $\phi$ and $X$ denote partial differentiation
with respect to these independent variables. It is clear that it is
the dependence of $G$ on the field's gradient, $X$, that prevents
this theory from being recast as a k-\emph{essence} model, whereas
the derivative of $G$ with respect to $\phi$ can be recast as an
additional k-\emph{essence} term. We will therefore only consider
such non-trivial models where $G_{X}\neq0$. The above form of the
Lagrangian also implies that no quantities will ever be dependent
on the function $G\left(\phi,X\right)$ itself, but only on its derivatives. 

\subsection{Equation of Motion for the Scalar Field}
The equation of motion for the scalar field can be obtained in the
usual way by varying the action
\begin{align*}
\frac{1}{\sqrt{-g}}\frac{\delta S_{\phi}}{\delta\phi}=\mathcal{P_{\phi}}-\nabla_{\mu}\left(\left(\mathcal{L}_{X}-G_{\phi}\right)\nabla^{\mu}\phi-\nabla^{\mu}G\right)\,,
\end{align*}
where $\mathcal{P}_{\phi}=K_{\phi}-\left[\left(\nabla^{\lambda}\phi\right)\nabla_{\lambda}\right]G_{\phi}$.
In particular, this presentation of the result makes explicit the
existence of a Noether current for Lagrangians invariant under constant
shifts of the field, $\phi\rightarrow\phi+c$,
\begin{equation}
J_{\mu}=\left(\mathcal{L}_{X}-2G_{\phi}\right)\nabla_{\mu}\phi-G_{X}\nabla_{\mu}X\,.\label{e:Jmu}
\end{equation}
In terms of this current, the equation of motion takes the 
elegant form 
\begin{equation}
\nabla_{\mu}J^{\mu}=\mathcal{P_{\phi}}\,.\label{eq:DivJeom}
\end{equation}
The fully expanded equation of motion is somewhat unwieldy, but its
structure is important for future discussion:
\begin{equation}
L^{\mu\nu}\nabla_{\mu}\nabla_{\nu}\phi+\left(\nabla_{\alpha}\nabla_{\beta}\phi\right)Q^{\alpha\beta\mu\nu}\left(\nabla_{\mu}\nabla_{\nu}\phi\right)+\Ecal_\phi-G_{X}R^{\mu\nu}\nabla_{\mu}\phi\nabla_{\nu}\phi=0\label{e:phi}
\end{equation}
where $R_{\mu\nu}$ is the Ricci tensor and $\Ecal$, $L_{\mu\nu}$ and
$Q_{\alpha\beta\mu\nu}$ are constructed out of the metric $g_{\mu\nu}$,
field $\phi$ and its first derivatives $\nabla_{\mu}\phi$ only.
The term $\Ecal_\phi$ which is zero order in the second derivatives is the field derivative of 
\begin{equation}
\label{EcalFirst}
\Ecal=2X\left(K_{X}-G_{\phi}\right)-K\,,
\end{equation}
the terms linear in the second derivatives are contracted with
 \[
L_{\mu\nu}=\left(K_{X}-2G_{\phi}+2XG_{X\phi}\right)g_{\mu\nu}+\left(K_{XX}-2G_{\phi X}\right)\nabla_{\mu}\phi\nabla_{\nu}\phi\,,
\]
and the terms quadratic in the second derivatives are contracted with
\begin{equation}
\label{Qabmn}
Q^{\alpha\beta\mu\nu}=\frac{1}{2}\left(g^{\alpha\beta}H^{\mu\nu}+H^{\alpha\beta}g^{\mu\nu}\right)-\frac{1}{4}\left(g^{\mu\beta}H^{\nu\alpha}+g^{\nu\beta}H^{\mu\alpha}+g^{\mu\alpha}H^{\nu\beta}+g^{\nu\alpha}H^{\mu\beta}\right)\,,
\end{equation}
where 
\begin{equation}
\label{Hmn}
H_{\mu\nu}=G_{X}g_{\mu\nu}+G_{XX}\nabla_{\mu}\phi\nabla_{\nu}\phi\,.
\end{equation}
Most importantly, the equation of motion contains at most second-order
derivatives, and therefore the theories with \emph{kinetic braiding}
do not contain any hidden degrees of freedom. Moreover, the structure
of the tensor $Q^{\alpha\beta\mu\nu}$ implies that $\ddot{\phi}$ appears only linearly in the equation of motion and therefore the equation of motion is solvable with respect to $\ddot{\phi}$, for any time coordinate\footnote{Lorentz symmetry implies that this statement is true for the second derivatives $\phi_{xx}$ with respect to any coordinate $x$%
}. This is sufficient to ensure that this equation
is of the normal type required by the Cauchy-Kowalewski theorem and that the Cauchy problem has a unique local solution, at least for analytic functions. 

The presence of a third-order derivative might have been expected owing
to the d'Alamber-tian term in the Lagrangian; however, these terms arise in the
combination $\left(\nabla\Box-\Box\nabla\right)\phi$ and can be commuted
away, leaving behind a term coupling the scalar to the Ricci tensor,
$R_{\mu\nu}$. This is the key feature of theories with \emph{kinetic braiding}. Namely a seemingly minimally coupled theory has an 
equation of motion depending explicitly on the curvature so that the
scalar equation of motion includes the second derivatives of both
the scalar field and the metric.

It is interesting to note that Eq.~\eqref{e:phi} can be considered as a generalisation of the Amp\`{e}re-Monge equation for four-dimensional manifolds with Lorentzian signature.

\subsection{Energy-Momentum Tensor}\label{s:EMT}
The energy-momentum tensor (``EMT'') for the scalar field is most
easily derived in the standard way using the Lagrangian presented
in Eq.$\,$\eqref{e:L2}
\begin{equation}
T_{\mu\nu}\equiv\frac{2}{\sqrt{-g}}\frac{\delta S_{\phi}}{\delta g^{\mu\nu}}=\mathcal{L}_{X}\nabla_{\mu}\phi\nabla_{\nu}\phi-g_{\mu\nu}\mathcal{P}-\nabla_{\mu}G\nabla_{\nu}\phi-\nabla_{\nu}G\nabla_{\mu}\phi\,.\label{e:EMT}
\end{equation}
It is key to observe that the EMT also contains second derivatives
of $\phi$ appearing in $\mathcal{P}$, $\mathcal{L}_{X}$ and $\nabla_{\mu}G$.
Therefore both the Einstein's equations and the equation of
motion for the scalar field, Eq.~\eqref{e:phi}, contain second
derivatives of both the metric $g_{\mu\nu}$ and the scalar field
$\phi$, so that the system is not diagonal in second derivatives.
This kinetic mixing is essential and cannot be undone
through a conformal transformation. We shall refer to such a mixing as \emph{kinetic braiding} and show that it leads to a number of surprising properties for this system. 

In particular, for general configurations with timelike field derivatives the structure of the EMT is not that of a perfect fluid, contrary to minimally coupled k-\emph{essence} theories, including canonical scalar fields. Moreover, for some of such configurations, the EMT does not even have a timelike eigenvector. The imperfect nature of the fluid is directly related to \emph{kinetic braiding}.  In the paper \cite{Imperfect}, we discuss the corresponding imperfect-fluid picture in detail. For cosmological solutions, which are exactly homogeneous and isotropic, the EMT is forced to be of perfect-fluid type. However, cosmological solutions including small perturbations possess an EMT with small terms deviating from the perfect fluid. This changes the standard picture of cosmological perturbations where the intuition was gained from the perfect fluid case. 

Using the Einstein equations including the EMT contribution $T_{\mu\nu}^{\text{ext}}$
of fields different from $\phi$, \begin{equation}
R_{\mu\nu}-\frac{1}{2}g_{\mu\nu}R=T_{\mu\nu}+T_{\mu\nu}^{\text{ext}}\,,\label{EinsteinEquation}\end{equation}
we can eliminate the second derivatives of the metric in the scalar
field equation of motion Eq.$\,$\eqref{e:phi}. In this way we obtain
\begin{align}
\tilde{L}^{\mu\nu}\nabla_{\mu}\nabla_{\nu}\phi+\left(\nabla_{\alpha}\nabla_{\beta}\phi\right)Q^{\alpha\beta\mu\nu}\left(\nabla_{\mu}\nabla_{\nu}\phi\right)+Z-\qquad\qquad\qquad\qquad\label{e:eomsc}\\
-G_{X}\left(T_{\text{ext}}^{\mu\nu}-\frac{1}{2}g^{\mu\nu}T_{\text{ext}}\right)\nabla_{\mu}\phi\nabla_{\nu}\phi=0\,,\nonumber \end{align}
where the new term which does not contain second derivatives is \[
Z=\Ecal_{\phi}-2XG_{X}\left(X\left(K_{X}-4G_{\phi}\right)+K\right)\,,\]
while the terms linear in the second derivatives are now contracted
with
\begin{equation}
\label{Ltilde}
\tilde{L}_{\mu\nu}=L_{\mu\nu}-2XG_{X}^{2}\left(Xg_{\mu\nu}-2\nabla_{\mu}\phi\nabla_{\nu}\phi\right)\,.
\end{equation}
The tensor $Q^{\alpha\beta\mu\nu}$ remains the same. Note
that the new contributions modifying $\Ecal$ and $L_{\mu\nu}$ are 
suppressed by $M_{\text{Pl}}^{2}$. In Eq.~\eqref{e:eomsc} we have obtained a form of equations of motion which contains the
second derivatives only of  the scalar field $\phi$ but not of the
metric%
\footnote{Here we assume that, as usual, other matter fields do not contain second
derivatives in their EMT.%
}. This is important for the investigation of causality and stability
in this class of models, see section~\S\ref{ss:metric}. 

Let us finally emphasise the fact that the scalar inevitably becomes
coupled to external matter in the non-linear fashion resulting from the last term of Eq.~\eqref{e:eomsc}. This occurs in a quite peculiar manner: unless an additional explicit direct scalar/matter interaction is introduced, this coupling is one way and external matter does \emph{not} become coupled to the scalar, only feeling it through its gravitational effects. This coupling will play a central role in the following sections, is a consequence of the \emph{kinetic braiding}. A proper analysis of the constraints on the model arising as a result of this coupling will be deferred to future investigation, but for some headway in this direction see Ref.~\cite{DeFelice:2010gb}. Let us only remark here that the self-interactions of the scalar can give rise to a Vainshtein mechanism that helps to evade the observational constraints.
\subsection{Effective Metric}
\label{ss:metric}
Let us perturb the system comprising the scalar equation of motion
and Einstein equations around a given solution $\phi\left(x^{\mu}\right)$
and $g_{\alpha\beta}\left(x^{\mu}\right)$. Without loss of generality,
we will neglect the perturbations in the external fields, so that $\delta T_{\mu\nu}^{\text{ext}}=0$.
Now let us find the characteristic surfaces $\mathcal{S}$ (wave-fronts)
for the linearized system. This is crucial for the investigation of
causality and stability with respect to the high-frequency perturbations,
see e.g. \cite{Courant}. 

The linearized equation of motion for the scalar field Eq.\ \eqref{e:eomsc}
does not contain second derivatives of the metric. Therefore the matrix
of the principal part of the differential operator for this system
is lower triangular in $\left(\delta\phi,\delta g_{\mu\nu}\right)$
coordinates. This implies that the characteristic equation for the
system (see e.g.\ \cite[p.\ 580]{Courant}) becomes just a product
of a purely gravitational and a purely scalar part and the second
derivatives of the scalar field still present in the linearised Einstein
equations do not play any role for characteristics. The causal structure for the propagation of the scalar perturbations can therefore be directly read from just
equation~\eqref{e:eomsc}. Note that the same procedure for the original equation of motion \eqref{e:phi} would give an incorrect result. 

From the standard eikonal ansatz $\delta\phi=\mathcal{A}\left(x\right)\exp\left(i\omega\mathcal{S}\left(x\right)\right)$
with a slowly varying amplitude $\mathcal{A}\left(x\right)$, taking
the formal limit $\omega\rightarrow\infty$, we obtain
\begin{equation}
\mathcal{G}^{\mu\nu}\partial_{\mu}\mathcal{S}\partial_{\nu}\mathcal{S}=0\,,
\end{equation}
where the effective \emph{contravariant} metric for propagation of
perturbations is
\begin{equation}
\mathcal{G}^{\mu\nu} \equiv \tilde{L}^{\mu\nu}+2Q^{\alpha\beta\mu\nu}\nabla_{\alpha}\nabla_{\beta}\phi\,.
\end{equation}
Using formulae \eqref{Qabmn} and \eqref{Ltilde}, we can expand
this expression as 
\begin{equation}
\label{metric}
\mathcal{G}_{\mu\nu}=\Omega g_{\mu\nu}+\Theta\nabla_{\mu}\phi\nabla_{\nu}\phi-\nabla_{\mu}\left(G_{X}\nabla_{\nu}\phi\right)-\nabla_{\nu}\left(G_{X}\nabla_{\mu}\phi\right)\,,
\end{equation}
where we have introduced the notation
\begin{equation}
\Theta\equiv\mathcal{L}_{XX}+4XG_{X}^{2}\,,
\end{equation}
and 
\begin{equation}
\Omega\equiv\mathcal{L}_{X}-2G_{\phi}+\nabla_{\lambda}\left(G_{X}\nabla^{\lambda}\phi\right)-2X^{2}G_{X}^{2}\,.\label{Omega}
\end{equation}
The velocity of the wavefront \emph{$\mathcal{S}$} and the corresponding
causal structure of the \emph{acoustic} spacetime should be inferred
from the inverse or \emph{covariant} metric $\mathcal{G}_{\mu\nu}^{-1}$
for which $\mathcal{G}_{\mu\lambda}^{-1}\mathcal{G}^{\lambda\nu}=\delta_{\mu}^{\nu}$,
for details see e.g.\ Appendix A of Ref.~\cite{Babichev:2007dw}. 
The effective metric for the decoupling limit of the DGP model (where $G\left(\phi,X\right) \propto X$) was first obtained in Ref.~\cite{Nicolis:2004qq}, see also e.g.\ \cite{Creminelli:2010ba}. The result \eqref{metric} is more general, because it is derived for a generic function  $G\left(\phi,X\right)$. 

The main obstacle for the analysis of this metric for general backgrounds is that $\mathcal{G}_{\mu\nu}$ contains not only second derivatives in $\Omega$ and $\Theta$ but also an additional tensorial structure which is a Lie derivative of the metric $\pounds_{V}g_{\mu\nu}=\nabla_{\mu}\left(G_{X}\nabla_{\nu}\phi\right)+\nabla_{\nu}\left(G_{X}\nabla_{\mu}\phi\right)\ $ with respect to $V_{\mu}=G_X\nabla_{\mu}\phi$. This structure prevents $\nabla_{\mu}\phi$ from being an eigenvector of the metric for a general background as was the case for k-\emph{essence}. 

In this paper, we will restrict our analysis to the cosmological case. In particular, in section \S\ref{stability}, we calculate the sound speed and present the condition for the absence of ghosts using the metric \eqref{metric}. The same results are confirmed by a direct calculation of the action for cosmological perturbations, presented in Appendix ~\ref{sec:apppert}. The advantage of this effective-metric formalism is that it allows for the study of causality and stability for high-frequency perturbations for any given background without calculating the action for perturbations, which may be a very complicated task for general, less symmetric backgrounds.    
\section{Cosmology}
\label{s:cosmo}
\subsection{Background Evolution}

In this section, we will discuss cosmology in braided models  \eqref{L}. 
We will first consider the homogeneous and isotropic background solutions $\phi(t)$ for a generic model and, in Section~\ref{s:shift}, we will specialize to the shift-symmetric case, where the integration of the equations of motion is almost straightforward.

For simplicity, we will restrict our discussion to the spatially flat Friedman-Robertson-Walker (``FRW'') metric 
\begin{equation}
    \ds s^2 = \ds t^2 - a^2(t)\ds \xb^2 \,.
\end{equation}
On this background, we have $X=\frac{1}{2}\dot\phi^2$,  where the dot represents the time derivative. In the current Eq.~\eqref{e:Jmu} only the ``charge density'' 
\begin{equation}
  J \equiv J_0= \left( K_X - 2G_\phi + 3H\dot\phi G_X\right)\dot\phi \,, \label{e:J}
\end{equation}
survives, so that the equation of motion \eqref{eq:DivJeom} takes the form 
 \begin{equation}
  \dot{J}+3HJ = \Pcal_\phi \,, \label{e:Jdot}
\end{equation}
where, as usual, $H\equiv \dot a/ a$ denotes the Hubble parameter. Note that this equation of motion not only contains $\ddot{\phi}$  but also $\ddot{a}$. As a result of the symmetry of the cosmological setup, the energy-momentum tensor takes the perfect-fluid form and the only non-vanishing stress-tensor components, Eqs~\eqref{e:EMT}, reduce to the energy density 
\begin{equation}\label{EnergyDensity}
  \epsilon \equiv T^{0}_{\phantom{i}0} = \dot\phi J + 2XG_\phi - K \,, 
\end{equation}
and the pressure   
\begin{equation}
\label{pressure}
   \Pcal  \equiv  - \frac{1}{3} T^{i}_{\phantom{i}i} =  K - 2XG_{\phi} -2XG_{X} \ddot{\phi}\,,
  \end{equation}
 which turns out to be given by the Lagrangian \eqref{e:L2}. The two unusual properties arising as a result of \emph{kinetic braiding} are that the pressure contains $\ddot{\phi}$  while the energy density depends on, $H$, the Hubble parameter\footnote{Note that this situation is different from the so-called \emph{Inhomogeneous Equation of State} where pressure is postulated to be a function of both the energy density and $H$ \cite{Nojiri:2005sr}.}. 
 
It is convenient to introduce the variable ``measuring'' the strength of the \emph{kinetic braiding} 
\begin{equation}
\label{kappa}
\kappa = XG_X\,;
\end{equation}
its physical meaning will be further elucidated in the paper \cite{Imperfect}. 

The part of energy density \eqref{EnergyDensity} which is \emph{not braided} and which does not depend on the Hubble parameter  looks similar to k-\emph{essence}:  
\begin{equation}
\label{Ecall}
\Ecal = 2X(K_X-G_\phi) - K \,,
\end{equation}
and its field derivative appears already in the equation of motion \eqref{e:phi}, so that 
\begin{equation}
\epsilon(\phi, \dot\phi, H)= \Ecal + 6 \kappa  \dot \phi H\,. \label{energyFunction}
\end{equation} 
The Friedmann equation becomes not the usual complete square but a general quadratic expression for $H$, 
\begin{equation}
 \label{e:H2}
    H^2 = 2\kappa \dot\phi H+\frac{1}{3}\left(\Ecal + \rho_\ext \right) \,,
\end{equation}
where we have introduced the energy density $\rho_\ext$ from any additional sector external to $\phi$. This could be dust, radiation, a bare cosmological constant term or any other standard cosmological fluid. 

It is useful to compare this equation with the Friedmann equation arising in the DGP scenario \cite{Deffayet:2000uy,Deffayet:2001pu}. In particular, we see that $(\kappa \dot \phi)^{-1}$ plays the role of the crossover scale $r_0$ beyond which gravity starts probing the extra dimension in DGP, while $\Ecal$ corresponds to the energy density localized on the brane. 

We can solve for $H$ in Eq.~\eqref{e:H2} to obtain
\begin{equation}
    H = \kappa \dot\phi +\sigma\sqrt{( \kappa \dot\phi )^2 +\frac{1}{3}\left(\Ecal + \rho_\ext \right) }\,, \label{e:H}
\end{equation}
with $\sigma=\pm 1$ denoting a choice of branch. 
We will see in section~\S\ref{s:stability} that only one choice of $\sigma$ (which depends on the forms of $K$, $G$) is compatible with stable fluctuations around the solution.

The presence of the term linear in $H$ in \eqref{e:H2} is responsible for a mismatch between the two branches of solutions (labelled by $\sigma$) and expansion or contraction of the cosmologies. Only a simultaneous transformation $(\sigma=+1,\dot\phi)\rightarrow(\sigma=-1,-\dot\phi)$ gives the time-reserved solution. 

For future reference, it is useful to also write down the acceleration equation,
\begin{equation}
 \label{e:Hdot}
    \dot{H} =-\frac{1}{2} (\epsilon+\rho_\ext+\Pcal+p_\ext) = \kappa \ddot\phi  -\frac{1}{2} J\dot\phi -\frac{1}{2} (\rho_\ext+p_\ext)\,, 
\end{equation}
where $p_\ext$ is the pressure of the external matter; we will also sometimes use its equation-of-state parameter, $w_\ext \equiv p_\ext/\rho_\ext$. \\

Finally we present the expanded form of the equation of motion \eqref{e:eomsc} in a cosmological context
\begin{equation}
  D\ddot\phi + 3J(H-\kappa \dot\phi) +\epsilon_\phi = 3\kappa(\rho_\ext+p_\ext)\,, \label{e:ddphi}
\end{equation}
where 
\begin{equation}
  D = \Ecal_X+ 6\dot\phi H\kappa_X+ 6\kappa^2 \,. \label{e:D} 
\end{equation}
In the absence of \emph{kinetic braiding}, $\kappa=0$, this reduces to the k-\emph{essence} equation of motion, c.f.\ Ref.~\cite[Eq.\ (II.13)]{Vikman:2004dc}. 
One notices that the external matter content explicitly filtered down into Eq.~\eqref{e:ddphi} despite the absence of any direct coupling between $\phi$ and matter in the Lagrangian. This is traced back to the explicit dependence of $J$ on $H$ and is a consequence of the braiding between the scalar and gravity in these models. 
The normalization factor $D$ in Eq.~\eqref{e:D} will appear repeatedly below and it plays a prominent role since it determines whether the fluctuations of $\phi$ around the background are ghosty ($D<0$) or not ($D>0$), see section~\S\ref{stability}. 

Notice the presence of a term quadratic in $\kappa$. Once the $M_{\text{Pl}}$ factors are restored, we see that this term is suppressed by one more power of $M_{\text{Pl}}^2$, reflecting that it is a type of backreaction effect implied by the scalar-gravity \emph{braiding}. This term arises from the dependence of the equation of motion for $\phi$, Eq.~\eqref{e:phi}, on $H$ and the simultaneous dependence of $H$ on $\dot\phi$ as implied by the Friedmann equation.

In the absence of external matter, close to the point where $J=0$, $\ddot{\phi}$ will be small provided $\epsilon$ is a slowly-varying function of $\phi$, through Eq.~\eqref{e:ddphi}. Then Eq.~\eqref{e:Hdot} implies that we are dealing with a generalised slow-roll, quasi-de-Sitter state.

\subsection{Stability}
\label{stability}
Now let us apply the machinery of the effective metric to cosmological solutions. Plugging in the cosmological background into the general expression for the metric \eqref{metric} we obtain for the \emph{contravariant} metric
 \begin{equation}
\mathcal{G}^{\mu\nu}=\delta_{0}^{\mu}\delta_{0}^{\nu}D-\delta_{i}^{\mu}\delta_{i}^{\nu}a^{-2}\left(\Omega-\frac{4\kappa}{\dot\phi}H\right)\,.
 \end{equation}
From this expression, it follows that ghosts are absent if $D>0$. Indeed, it is $\mathcal{G}^{\mu\nu}$ which is contracting the derivatives in the kinetic term for perturbations\footnote{In the high wave-number limit, the gauge-invariant canonical variable $v$ is proportional to the gauge invariant $\delta \phi$. Also note that the full calculation of the quadratic action gives the expression for the normalisation of the cosmological perturbations, Eq.~\eqref{A_app}. This is modified compared to the standard case and therefore is likely impact predictions of inflation.}, see \eqref{quadraticaction}.  
Here it is important that the sign in front of $\mathcal{G}^{\mu\nu}$
be chosen correctly---so that this condition continuously transforms
to that of k-\emph{essence}. 

This  \emph{contravariant} metric is diagonal. Therefore it is easy to find its inverse (or \emph{covariant}) metric along with the corresponding \emph{acoustic} line element
\begin{equation}
\label{interval}
\mbox{d}S^{2}=\mathcal{G}_{\mu\nu}^{-1}\mbox{d}x^{\mu}\mbox{d}x^{\nu}=\left(\Omega-\frac{4\kappa}{\dot\phi}H\right)^{-1}\left(c_{\ssl}^2\mbox{d}t^{2}-a^{2}\left(t\right)\mbox{d}\xb^{2}\right),
\end{equation}
where the sound speed is given by 
\begin{equation}
c_{\ssl}^{2}=\frac{\Omega\dot\phi-4\kappa H}{\dot\phi D}\,.
\label{SoundSpeedFRW}
\end{equation}
Thus for the stability with respect to high-frequency perturbations
we should require that\footnote{Note that the tensor modes are not affected in our model, as can be seen in the full calculation of the quadratic action Eq.~\eqref{quadraticaction}, and therefore do not provide any new conditions for stability.}
\begin{eqnarray}
D>0 & \mbox{and} & \Omega-\frac{4\kappa}{\dot\phi}H>0\,.\label{StabilityConditionsFRW}
\end{eqnarray}
Only those models for which during the whole evolution history these two conditions are satisfied can be considered as plausible from a physical point of view. Further we calculate $\Omega$ for the cosmological background and obtain
\begin{equation}
\Omega-\frac{4\kappa}{\dot\phi}H=\frac{J+2\left(\dot{\kappa}+H\kappa-\kappa^{2}\dot{\phi}\right)}{\dot{\phi}}\,,
\end{equation}
while the definition \eqref{e:D} can be rewritten as $D\dot{\phi}=\epsilon_{\dot{\phi}}-6H\kappa+6\kappa^{2}\dot{\phi}$ where the partial derivative of the energy density (which is considered as a function of three independent variables, see Eq.~\eqref{energyFunction})  is taken keeping $H=\text{const}$, as if the Hubble parameter were externally fixed. 

Using these expressions, the formula for the sound speed \eqref{SoundSpeedFRW} can be written as 
\begin{equation}
c_{\ssl}^{2}=\frac{J+2\dot{\kappa}+2\kappa\left(H-\kappa \dot \phi\right)}{\epsilon_{\dot\phi}-6\kappa\left(H-\kappa \dot\phi\right)}\,.
\end{equation} 
This expression depends on $\ddot\phi$ which can be eliminated using the equation of motion \eqref{e:ddphi}. 
It is important to note that
\begin{equation}
c_{\ssl}^{2}=\frac{\mathcal{P}_{\dot{\phi}}+4\dot{\kappa}+2\kappa\left(4H-\kappa\dot{\phi}\right)}{\epsilon_{\dot{\phi}}-6\kappa\left(H-\kappa\dot{\phi}\right)} \neq\frac{\dot{\mathcal{P}}}{\dot{\epsilon}}\,,
\end{equation}
even in the shift-symmetric case, contrary to k-\emph{essence} \cite{Garriga:1999vw}. Here the derivative of the total pressure is taken keeping $\ddot\phi=\text{const}$.  The sound speed in these models is arbitrary and can even be larger than the speed of light. In particular, this is exactly the case in the model discussed in the section~\S\ref{s:linmodel}, see e.g.\ Fig.~\ref{f:attrevol}.  In this respect the situation is not very different from  k-\emph{essence}\footnote{However, contrary to the case of k-\emph{essence} and the conclusions of Ref.~\cite{Dubovsky:2005xd}, in the model presented in  section~\S\ref{s:linmodel} there is a period during which the null-energy condition is violated on an isotropic background while the speed of sound is sublumninal, see Figs~\ref{f:attrevol} or \ref{f:Jdom}.}. From the expression for the acoustic interval Eq.~\eqref{interval}, it follows that there are no causal pathologies (Closed Causal Curves) even in the presence of superluminal propagation.  
We follow the discussion of this issue presented in Ref.~\cite{Babichev:2007dw} (see also the most recent paper \cite{Geroch:2010da} and older works \cite{ArmendarizPicon:2005nz, Bruneton:2006gf, Bruneton:2007si,  Kang:2007vs}) and claim that this superluminality does not imply any inconsistencies. 
A different issue is a possible UV completion of the theories with superluminal propagation. Currently, there are good arguments \cite{Adams:2006sv} that this UV completion cannot be realized within a Lorentz-invariant and renormalizable local QFT or perturbative string theory framework. However, on the level of an effective-field-theory description there are no inconsistencies. This is also accepted by some of the authors of \cite{Adams:2006sv}, see \cite{Creminelli:2010ba, Dubovsky:2005xd}. This possibility to realise superluminal propagation of perturbations on nontrivial backgrounds can lead to interesting cosmological scenarios in the context of inflation \cite{Mukhanov:2005bu} and some exotic alternatives, e.g.\ \cite{Creminelli:2010ba, Bessada:2009ns, Magueijo:2010zc, Magueijo:2008sx}. 
Moreover, this superluminality raises important questions regarding black hole physics \cite{Dubovsky:2006vk,Babichev:2006vx,Babichev:2007wg,Eling:2007qd}. 

\section{Cosmology of Shift-Symmetric Gravity-Braided Models}
\label{s:shift}

In this section, we will restrict our attention to models that realise an exact shift symmetry 
$$
\phi\to\phi+\const\,,
$$
in terms of an appropriate choice of the scalar-field variable $\phi$. In practice, this implies that $K,G$ must be $\phi-$independent, that is the scalar Lagrangian takes the form\footnote{Shift-invariance allows for somewhat more freedom in choosing Lagrangians. Obviously, a term in $G$ linear in  $\phi$ is compatible with this symmetry. However, it can be absorbed in the form of $K$ by integration by parts. Similarly, a change of variable $\phi\to \tilde\phi(\phi)$ will trivially introduce $\tilde\phi$-dependence but will still realize the same symmetry, even though not as a shift in $\tilde\phi$.}
\be\label{Lshift}
	\Lcal=K(X)+G(X)~\Box\phi\,.
\ee
Imposing such a symmetry, i.e.\ requiring that the field only be derivatively coupled, means that $\phi$ can be interpreted as a Goldstone boson of some broken symmetry. This naturally prevents such a field from acquiring a mass. As we will see, the dynamics of such kinetically braided Goldstone bosons make them a very compelling model for dark-energy dynamics. In addition, 
the scalar would have to be  derivatively coupled to matter, if at all, which would make evading fifth-force constraints relatively easy\footnote{Generic derivative couplings to matter can still give rise to sizeable effects, but typically the most problematic couplings can be suppressed by additional symmetries \cite{ArkaniHamed:2003uy}. In fact, in this sense the situation in the present model quite similar to  ghost condensation \cite{ArkaniHamed:2003uy} and Ho\v rava gravity \cite{horava,bps2,Blas:2010hb}}.

In the shift-symmetric case, the shift-current \eqref{e:Jmu} is conserved, and the scalar equation of motion is precisely this statement, $\nabla_\mu J^{\mu}=0$.
In a homogeneous and isotropic background, this reduces to simply
\begin{equation}
   \dot{J} + 3HJ =0 \,, \label{e:Jdot_shift}
\end{equation}
that is, the shift charge in a comoving volume is constant. 
Furthermore, the equation of motion Eq.~\eqref{e:Jdot_shift} can be trivially integrated, 
\begin{equation}
  J = \dot\phi(K_{X} + 3\dot\phi H G_{X}) = \frac{\const}{a^3(t)} \,. \label{e:Jattr}
\end{equation}
Hence, the expansion of the universe drives $J$ to zero, and the locus of $J=0$ in configuration space represents a future attractor for expanding FRW space-times\footnote{The appearance of de-Sitter attractors in shift-symmetric theories with higher derivatives was also used in the so-called the B-Inflation \cite{Anisimov:2005ne}. This behaviour was first noticed by Alexei Anisimov, one of the authors of the B-Inflation paper, who regrettably passed away after this preprint was submitted to the arXiv, RIP.}. As we will see, even on the attractor one can have rather interesting behaviour, basically because the value of $\dot\phi$ on the attractor depends on the amount of external matter density at every moment of time (because of the appearance of $H$ in $J$). 

It is convenient and illuminating to split the energy density of the scalar into the contribution from the attractor and that of the departure from it,
\begin{align}\label{split}
&\epsilon = \epsilon^* + \epsilon^J~,\cr
&{\rm with} \quad \epsilon^*\equiv\epsilon|_{J=0} \cr
&{\rm and} \quad \epsilon^J \equiv \epsilon-\epsilon^*~.
\end{align}
Both `components' behave like fluids with a particular equation of state on the background. By definition, the $J$ (or `off-attractor') component dilutes away with expansion, so it behaves like a rather standard fluid. As we will see, the attractor component can be quite exotic, as it can easily exhibit phantom behaviour without generating instabilities. In fact, we will find that requiring that the energy density in the attractor be positive and that the field fluctuation be healthy (not a ghost) implies that generically the attractor displays phantom behaviour. Hence, the composition of the $J$- and attractor components (that is, the situation with a generic initial condition) is such that the energy density in  $\phi$ first dilutes away and then grows. That is, we will find that $\phi$ can cross the phantom divide without leading to instabilities.

At this point, we note that since the action contains second derivatives, therefore possibility of smoothly crossing of the $w_X=-1$ barrier does not contradict the statement proved in \cite{Vikman:2004dc} and rederived in different ways later in \cite{Caldwell:2005ai, Hu:2004kh, Zhao:2005vj, Sen:2005ra, Abramo:2005be, Kunz:2006wc, Babichev:2007dw}. Therefore \emph{kinetic braiding} provides a working example of the so-called Quintom scenario of Ref.~\cite{Feng:2004ad}, see also reviews \cite{Cai:2009zp, Zhang:2009dw, Copeland:2006wr}. Thus in this respect, \emph{kinetic braiding} exhibits similar phenomenology to models with explicit nonminimal coupling to gravity, which also allow one to have a classically stable crossing of the phantom divide in scalar-tensor theories \cite{Boisseau:2000pr, Gannouji:2006jm, Hu:2007nk, Motohashi:2010tb}. Another theory comprising a single degree of freedom which is able to penetrate the phantom divide without classical instabilities is the so-called $\lambda\varphi$-fluid (``Angel Dust'') \cite{Lim:2010yk}. However, this is a very non-standard theory where issues related to the strong-coupling scale and quantisation remain to be addressed. In the case of \emph{kinetic braiding}, the quantisation of perturbations is standard and the crossing occurs in a regime free of negative energy perturbations and which is completely under control. Models which exhibit healthy phantom behaviour within the Galileon framework were realised in Refs~\cite{Creminelli:2010ba,Nicolis:2009qm}. Moreover, in Ref.~\cite{Creminelli:2008wc} a healthy violation of the Null Energy Condition was achieved by including $\Box\phi\left(\partial\phi\right)^{2}$ on the level of perturbations within the effective-field-theory framework. 

A central issue in the significance of the split \eqref{split} is if and when the off-attractor component dilutes away sufficiently fast so that the attractor regime is actually reached. We discuss this among other points in \S\ref{s:approach}. But first, in the next subsection, we shall analyse the generic properties of attractor solutions.

\subsection{The Phantom Attractor}
\label{s:attractor}

First of all, since the Friedmann equation fixes $H$ as a function of $X=\dot\phi^2/2$ and the external matter energy density $\rho_{\ext}$, one can view $J=\dot\phi(K_{X} + 3\dot\phi H G_{X})$ as a function of the two variables $X$ and $\rho_{\ext}$,
$$
J=J(X,\rho_{\ext})~,
$$ 
whose form depends only on the choice of $K,G$. The attractor solutions are the roots of 
\beq\label{attr}
J(X_*,\rho_{\ext})=0
\eeq
We will furnish quantities evaluated on the attractor with an asterisk (`*') sub-/superscript\footnote{In principle, one should introduce a label to distinguish among the several possible attractors each of which has its own non-overlapping basin of attraction. In addition, the form of $J(X,\rho_{\ext})$ also depends on the branch for the Friedman equation ($\sigma$), so the attractors in general also depend on $\sigma$. 
For the sake of clarity, we will spare the reader these labels with no risk of introducing ambiguities.}.
In particular, for any form of external matter, the attractor behaviour of the scalar is such that $X$ becomes a certain local function of $\rho_{\ext}$, 
\beq\label{X*}
X_*=X_*(\rho_{\ext})~,
\eeq
the form of which is solely determined by the dynamics, i.e.\ by the form of $K,G$.
Notice that this holds irrespective of the time dependence of $\rho_{\ext}$ (i.e.\ its equation of state), so implicitly Eq.~\eqref{X*} also represents the time evolution of $X(t)$ on the attractor, $X(t)=X_*(\rho_{\ext}(t))$. 

From \eqref{EnergyDensity} it follows that on the attractor the energy density stored in the scalar, $\epsilon^*$, is also only a function of $\rho_{\ext}$,
\beq\label{e*}
\epsilon^* = \epsilon^*(\rho_{\ext}) \equiv -K(X_*(\rho_{\ext}))\,.
\eeq
Hence, the attractor has the remarkable property of responding to the external energy density in a way determined by the form of $K$ and $G$. 

We should emphasize that since $\epsilon^*$ is a function of $\rho_{\ext}$ only, this effectively allows one to  modify the Friedman equation (viewed as the relation between $H$ and $\rho_{\ext}$) in almost any way in these models. $K$ and $G$ can be picked in such a way so as to replicate any evolution history, once the shift current dilutes and the attractor is reached. The attractor present in our kinetically braided model can therefore be considered as a concrete realisation of the phenomenological ``Cardassian'' scenario proposed in Ref.~\cite{Freese:2002sq}. 

Of course, one must make sure that the perturbations are healthy, which will not be the case for a generic choice. It should be noted that these modifications are the result of the fact that scalar-gravity braiding modifies the Hamiltonian constraint in gravity.

The behaviour of the attractor can be easily understood in terms of the function 
$\epsilon^*(\rho_{\ext})$. For example, if $\epsilon^*(\rho_{\ext})$ is a decreasing  function (such as the one in the model of section~\S\ref{s:linmodel}), then as matter dilutes the attractor energy density grows, implying a phantom equation of state. In the presence of usual external matter ($w_\ext>-1$), $\epsilon^*$ will eventually dominate the expansion of the universe and, as long as $\epsilon^*(0)>0$, the final state will asymptote to de Sitter. We will see shortly that it is precisely this kind of setup, with phantom behaviour for the attractor, which is free of instabilities. 

At this stage, the form of the function $\epsilon^*(\rho_{\ext})$ is inevitably implicit---below we will elaborate on how to extract it in general once $K$ and $G$ are given. However, for most purposes its explicit form is not really necessary.

For instance, the effective equation of state of the attractor can be written as
\beq\label{w*}
1+ w_X^*  
=(1+w_{\ext})\frac{\ds \ln\epsilon^*}{ \ds \ln \rho_{\ext}} 
=(1+w_\ext)\rho_{\ext} \frac{ 6\kappa_*^2}{K_*D_*}
\eeq
The last equality follows from the chain rule and reading off $\frac{\ds X_*}{\ds \rho_{\ext}}=\frac{\dot\phi_*\ddot\phi_*}{\dot\rho_{\ext}}$ from the scalar equation of motion on the attractor,
\begin{equation}
   D_* \, \ddot\phi_* = - \kappa_* \frac{\dot\rho_\ext }{H} \,. \label{e:ddp_attr}
\end{equation}
Since the normalisation factor $D_*>0$, so that the $\phi$ fluctuation not be a ghost we find that (assuming $w_{\ext}>-1$):
\begin{itemize}
\item the attractor is phantom ($w^*_X<-1$) whenever its energy density $\epsilon^*=-K_*$ is positive, and vice-versa; 
\item as the external energy density dilutes away, the attractor approaches de Sitter with $w_X^*\rightarrow -1$ from below---the phantom region, provided $\epsilon^*>0$;
\item the equation of state of the attractor contribution is tied to the equation of state of the external matter and therefore will change, for example,  at matter-radiation equality.
\end{itemize}
It is quite simple to imagine that it is the behaviour of the attractor that is relevant for the observations of dark energy. Therefore, provided that the off-attractor contribution has diluted away early enough during the history of the universe, it is a generic prediction of the models of Eq.~\eqref{Lshift} that dark energy will be seen to violate the NEC and approach de Sitter from below---the phantom region.

Equations \eqref{e*} and \eqref{w*} encode the most important feature of the behaviour of the attractor, namely that $\phi$ responds to the presence of any other form of matter, and that a stable  quasi-de Sitter stage is reached only if the attractor is phantom. So far, we only used the positivity of the energy density ($-K_*$) and $D_*$. An additional constraint required for stability is discussed in \S\ref{s:stability}. \\

Let us close this subsection with some gymnastics that can ease the task of finding the attractor, in particular the function $\epsilon^*(\rho_{\ext})$, for an arbitrary $K$ and $G$. The technical difficulty is that the explicit expression of $J(X,\rho_{\ext})$ found from Eqs~\eqref{e:J} and \eqref{e:H} is rather involved and may not allow to solve easily for its roots. However, on the attractor, the equations simplify considerably---for example, the Friedman equation becomes
\begin{equation}\label{e:H2attr}
    3H^2_* = -K_* + \rho_\ext \,.
\end{equation}
From the equation $J=0$ itself we see that we can also write $H_* = - K^*_{X} / (3 \dot\phi_* G^*_X) $. 
Thus, it is easy to see that solving $J=0$ is equivalent to solving\footnote{Eq.~\eqref{e:attralt} needs to be supplemented with \eqref{e:sign} because the l.h.s. of Eq~\eqref{e:attralt} contains less information than $J$. Specifically, Eq~\eqref{e:attralt} does not involve $H$ and hence is independent of the choice of branch of the Friedman equation, $\sigma$. Eq.~\eqref{e:sign} provides that information.} 
\begin{align}
    & \frac{K_{X*}^2}{6X_*G_{X*}^2 } + K_* = \rho_{\ext}    \qquad \text{and} \label{e:attralt} \\[3mm]
	& \sgn\left(\frac{K_*-X_*K^*_{X}}{K^*_{X}}\right) = \sigma \sgn\dot\phi_* \,. \label{e:sign}
\end{align}
Clearly, this form of the equations for the attractor is considerably simpler. In particular, Eq.~\eqref{e:attralt} directly gives $\rho_{\ext}$ as a function of $X_*$, which is just the inverse of the function in which we are interested, $X_*(\rho_{\ext})$. Also, it is straightforward to obtain the relation between the attractor energy density $\epsilon^*$ and $\rho_{\ext}$ in parametric form,
$$
\Big(\epsilon^*(X),\,\rho_{\ext}(X) \Big)
=\Big( -K(X) \,,\; \frac{K_X(X)^2}{6X G_X(X)^2} + K(X) \Big)\,,
$$ 
in terms of the `parameter' $X$. This curve in the $\epsilon^*$--$\rho_{\ext}$ plane encodes how the attractor responds to the external matter $\rho_{\ext}$ for given $K,G$ (the $\sigma=\pm$ branches corresponding to consecutive segments of the curve).

Finally, let us mention that the roots of
\beq\label{dS}
{K_{X_*}^2 \over 6X_* G_{X*}^2} + K_* =0
\eeq
represent the pure-de-Sitter attractors (when $K_*<0$), corresponding to the asymptotic state of expanding solutions, once all matter and shift-current density, $J$, dilute away. These are kinetic condensates similar to ghost condensation \cite{ArkaniHamed:2003uy} in that the non-trivial $\dot\phi$ condensate mimics a cosmological constant. 
However, let us emphasize that the present situation is qualitatively different from ghost condensation for two reasons: First, in ghost condensation (the $G=0$ case) the attractor can only behave like a pure cosmological constant. Instead, with $G\neq0$ it has a nontrivial (phantom) equation of state in the presence of external matter. Second, the fluctuations around the attractor behave very differently in  ghost condensation and the model with \emph{kinetic braiding} \eqref{Lshift}. Even on the pure de Sitter condensate the sound speed squared of the scalar fluctuation can be positive, $c_\ssl^2>0$. This is a consequence of the presence of the of the $G\Box \phi$ term and the resulting imperfection of the effective fluid. As we will see in \S\ref{s:approach}, around the de Sitter condensate, the equation of state corresponding to a homogeneous perturbation (the off-attractor component due to $J\neq 0$) vanishes. Hence, the off-attractor component behaves like a pressureless fluid.  For a perfect fluid  this would imply that $c_\ssl^2=0$, as indeed happens in ghost condensation (which is why one needs to resort to higher space-derivative terms in Ref.~\cite{ArkaniHamed:2003uy}). In our case, despite having $w^J=0$ for the off-attractor component, one finds $c_\ssl^2>0$ (as long as  $H\neq0$). This can also be seen from the EFT point of view by including operators like $G \Box \phi$ in the Lagrangian for perturbations \cite{Creminelli:2008wc}.

\subsection{Approach to the Attractor}
\label{s:approach}

The background dynamics away from the attractor are given by Eq.~\eqref{e:Jattr}. 
The quantities that evolve with time in a simple way are $J \sim a^{-3} $ and $\rho_\ext \sim a^{-3(1+w_\ext)}$. In principle, then, to explicitly work out the time evolution of the system
we need to express everything in terms of $J$ and $\rho_\ext$. Unfortunately, this may not be possible even in concrete examples. Still, one can extract all the information from the  functional form of $J(X,\rho_\ext)$.
For simplicity let us concentrate on the $X$-dependence and let us consider a generic form of $J(X)$ such as depicted in Fig.~\ref{f:Jevol}.\footnote{The inclusion of nontrivial matter in the discussion is straightforward, it only adds a $\rho_\ext$ direction to the plot of Fig.~\ref{f:Jevol}.}  
The attractors so far discussed correspond to the cuts of $J$ with the $X$ axis. 

It is easy to check that the normalization factor for the perturbations, $D$, introduced in  Eq.~\eqref{e:D} in the shift symmetric case is nothing but
\beq\label{JxD}
D = \left( 1-\frac{\dot\phi \kappa}{H} \right) \frac{\partial J(\dot\phi,\rho_\ext)}{\partial \dot\phi}\Big|_{\rho_\ext = \const}\,.
\eeq
Therefore, we can readily identify non-ghosty regions in the plot of $J(\dot\phi)$ as those where
\begin{equation}\label{e:Dsigns}
	\sgn(D) = \sigma \sgn(H)  \sgn(J_{\dot\phi})=+1\,,
\end{equation}
Therefore, given a fixed choice of branch, $\sigma$, the perturbations will alternate between ghosty and healthy whenever $H$ or the slope $J_{\dot\phi}$ change sign. The roots of $J_{\dot\phi}=0$ separate the basins of attraction of the different attractors, representing edges between them where $D=0$. This implies the presence of a pressure singularity which the dynamics cannot penetrate and that the sound speed diverges ($c_\ssl^2 \simeq 1/D$ for $D\to0)$.

\EPSFIGURE{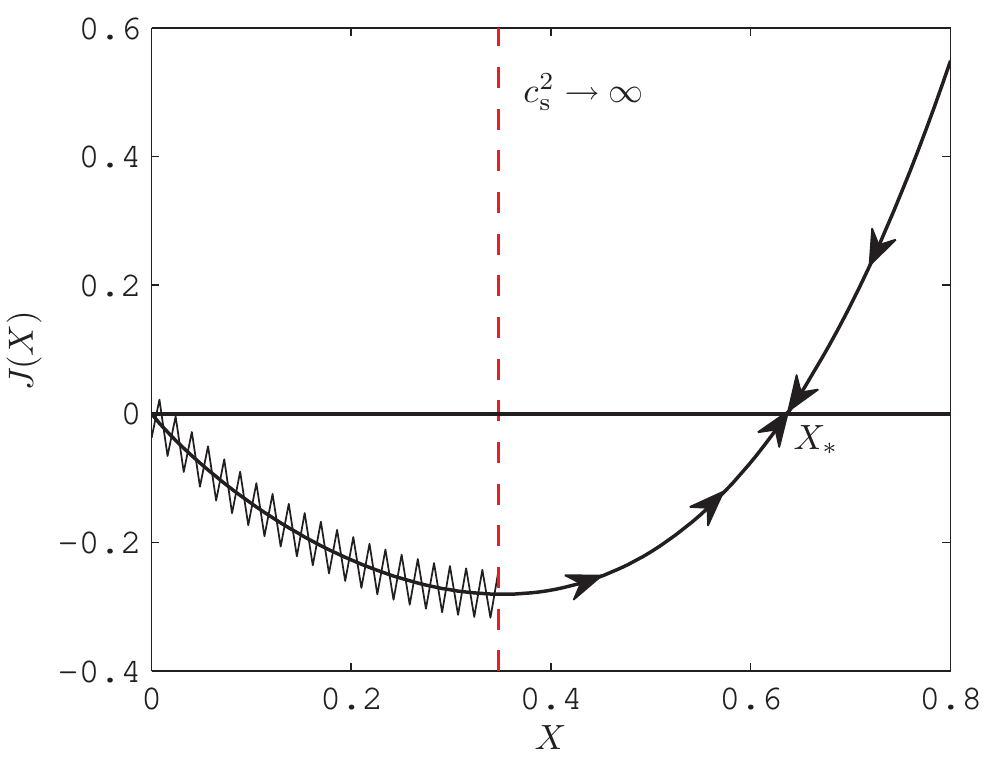}{\label{f:Jevol} The shift charge $J$ as a function of $X$ for an illustrative shift-symmetric model with kinetic braiding; we have suppressed the additional direction $\rho_\ext$. The loci of zeroes of $J$ are the attractors: evolution proceeds toward the locus marked as $X_*$. Typically, the attractors alternate between being stable and unstable. The basins of attraction are separated by a line of extrema of $J(X)$ (marked with the dashed vertical red line). There, the normalization of perturbations vanishes, $D = J_{\dot\phi}(H-\dot\phi \kappa)/H =0$. This leads to a pressure singularity, which the dynamics cannot penetrate, and a divergent sound speed, $c_\ssl^2\propto 1/D$ (see the discussion around Eq.~\eqref{SoundSpeedFRW}). The region on the left, marked with the jagged line, has the wrong sign of kinetic term, $D<0$ and therefore is does not represent a healthy background.}

In the following, we shall concentrate on the evolution close enough to a $J=0$ attractor. Our approximation requires that two conditions be met: firstly, $\epsilon^J/\epsilon^* \ll 1$, where we are using the splitting of the fluid introduced in Eq.~\eqref{split}; secondly, we must be close enough to the attractor such that we are away from any extrema in $J$, where the backwards-in-time evolution would meet a pressure singularity.

We will therefore study the equation of state of the fluid representing the excess over the attractor, with energy density $\epsilon^J = \epsilon-\epsilon^*$. Assuming that one can invert $J(X,\rho_\ext)$ to find $X(J,\rho_\ext)$, we can simply expand the total energy density
$\epsilon(J,\rho_\ext) = \dot\phi(J,\rho_\ext)J - K(X(J,\rho_\ext))$
for small $J$ to find, at lowest order in $J$,
$$
\epsilon^J \simeq \Xi \dot\phi_* \, J\,, \quad \text{where} \quad \Xi\equiv 1 - \frac{K^*_X}{D_*}\left(\frac{H-\dot\phi_* \kappa_*} {H}\right)\,.
$$
Notice that the factor in front of $J$ is a function of $\rho_\ext$, implying that $\epsilon^J$ does not dilute like dust as could have been naively expected and as is seen in the case of ghost condensates. Rather, the equation of state of the off-attractor component is found to be
\beq\label{wJ}
w^J=(1+w_\ext)\frac{K_X}{2D_*}\Omega_\ext\,\left( 1 + 2\left. { \partial \ln\Xi \over \partial\ln X}\right|_* \right)\,,
\eeq
where we have defined the contribution to total energy density of the external matter as $\Omega_\ext \equiv \rho_\ext/3H^2$. Hence, we see that the off-attractor contribution is also sensitive to the external matter, though differently from the attractor. In particular, we see that $J$ behaves like dust only when external matter is either absent or when it is just a cosmological constant.

It is interesting to note that, assuming the $\Xi$-derivative in the parentheses of Eq.~\eqref{wJ} is not negative enough to change the sign, the energy density stored in the current will redshift away more slowly than the external matter. This is irrelevant for the purpose of the discussion of the cosmological dynamics, since our approximation assumed that the energy density stored in the imperfect scalar will be dominated by its attractor part. However, it is true that the minimum of the ratio $\epsilon^J/\rho_\ext$ will be reached around the time of the transition from the $J$-dominated to attractor-dominated behaviour for the scalar fluid.

Finally, this discussion has only covered the case when $\epsilon^J$ is small (as compared to either $\epsilon^*$ or $\rho_\ext$). The opposite regime, when $\epsilon$ is dominated by the shift current will have significantly different behaviour. The nonlinearity of the equations make a general analysis somewhat difficult, so we defer an explicit analysis of this regime to  the example of \S\ref{s:linmodel}.
\subsection{Stability on Attractor}
\label{s:stability}

Specialising to the shift-symmetric case allows us to concentrate on the attractor, where significant simplification occurs since $J=0$. In particular, using Eq.~\eqref{e:ddp_attr}, the speed of sound can be expressed as
\begin{equation}
c_{\ssl*}^2 = \frac{2\kappa_*(H-\dot\phi_* \kappa_*)}{\dot\phi_*D_*} + \frac{ 6\kappa_*\kappa_{X*} (\rho_\ext + p_\ext)}{D_*^2}
\end{equation}
The second term results from the coupling of the scalar to the other matter content and will dilute away as the   scalar becomes the dominant source of energy density. When the scalar on the attractor dominates the energy density, we can express the sound speed positivity condition as
\begin{equation}\label{e:cs2dS}
 c_{\ssl*}^2 = \frac{ K_{X*}(K_*-X_*K_{X*})}{9H^2D_*} > 0
\end{equation}
which combined with Eq.~\eqref{e:sign} implies that only one of the branch of the Friedman equation, that for which $\sigma = \sgn \dot\phi_*$, is stable. This equation also displays another difference between this theory and ghost condensation: even on the pure-de Sitter condensates the sound speed in the braided model is nonzero, and $c_{\ssl*}^2$ can be adjusted to be positive. The observation that an operator like $G \box\phi$ can lead to $c_{\ssl*}^2>0$ in an expanding universe was also seen in \cite{Creminelli:2008wc} in the effective-field-theory language.  It should be noted that the linear model presented in section \S\ref{s:linmodel} does have a vanishing sound speed on the de-Sitter attractor once all the external matter dilutes away. In general, this will not be the case, however.

\section{Example: Imperfect Dark Energy}
\label{s:linmodel}

In this section, we will describe the properties of the cosmological solutions of the simplest model with \emph{kinetic braiding} which exhibits the interesting behaviour described up to this point. We will choose functions $K$ and $G$ to be shift symmetric and linear in $X$ and choose the background Minkowski space ($X=0$) to be ghosty in order to make the de-Sitter attractor the final and stable point of the evolution. The simplest such choice is
\begin{align}
  K &= -X \,,\label{e:IDE}\\
  G &= \mu X \,,\notag
\end{align}
where $\mu$ is a coefficient with mass dimension $-3$.

Beneath we will in turn discuss the phase space of this theory and the regions of interest, the behaviour of the attractor solution, where we will find that $w_X<-1$ while the sound speed squared remains positive. We will then detail how this attractor is approached and the behaviour of the energy density of the scalar during this time. Finally, we will discuss the regions of parameter space for this model which are not excluded by observations. We invite the reader to familiarise themselves with Fig.~\ref{f:phaseflat} and its caption: it provides an overview of the dynamics of the system which are described in detail in the rest of this section.

We have named this model \emph{Imperfect Dark Energy}. The background EMT for cosmological solution has, of course, a perfect-fluid form. However, the EMT for the solution including small perturbations can no longer be so described; the fluid is imperfect, as we have shown in section \S\ref{s:EMT}. It is exactly this modification of the fluid's properties which allows this model to circumvent certain no-go theorems and stably violate the Null-Energy Condition.

Other authors have also considered phenomenological models of dark energy which include deviations of the energy-momentum tensor away from the perfect-fluid form such as anisotropic stress \cite{Koivisto:2005mm, Mota:2007sz, Koivisto:2008ig}. 

\EPSFIGURE{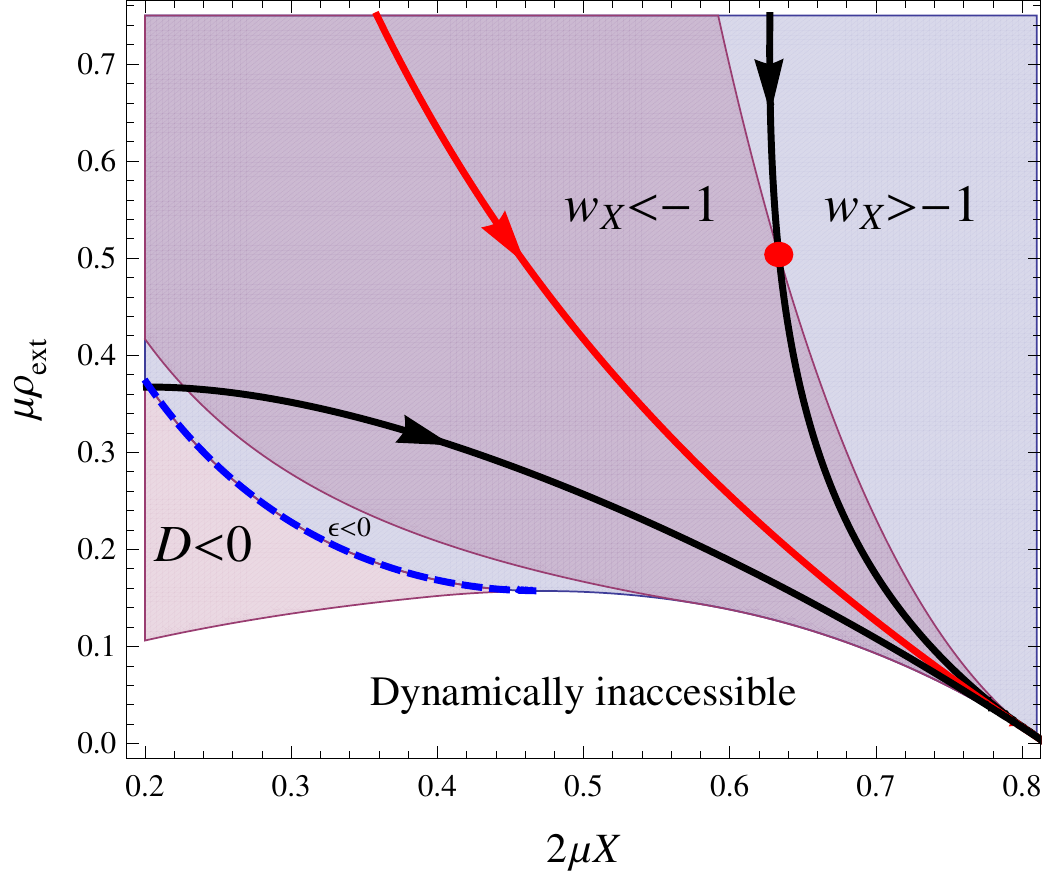,width=14cm}{\label{f:phaseflat} The phase space for the Imperfect Dark Energy model Eq.~\eqref{e:IDE} in the presence of external dust. The red line represents the dynamical attractor (where the shift charge $J=0$), with evolution occurring toward the final de-Sitter state in the bottom right. The inner purple-coloured region surrounding the attractor corresponds to phantom dynamics, $w_X<-1$. Outside it, on both sides, lie the lighter-coloured non-phantom regions, $w_X>-1$. The black lines depict example trajectories: starting with non-zero shift charge, the system evolves toward the attractor. Should the initial $J$ be large enough, the equation of state is non-phantom: the example trajectory on the right crosses the phantom divide at the position of the red blob. In this paper, we focus on the region to the right of the attractor since the energy density is always positive there. In the region to the left, the pressure is negative throughout and it is the fact that the energy density becomes negative in the region $\epsilon<0$ that makes the equation of state non-phantom. Further to the left lies the $D<0$ region separated by the dashed blue line on which $D$ vanishes and which is the locus of the minimum of $J$. This blue line is a pressure singularity and the region $D<0$ is not accessible dynamically from the healthy part of the phase space. In the whole region to the right of this dashed blue line, $c_\ssl^2$ is positive and perturbations have positive kinetic term. No solutions exist in the white ``dynamically inaccessible'' region.}

\subsection{Phase Space}
\label{s:linphase}
The phase-space plot for $\dot\phi$ for an FRW cosmology, extended to show $J$ and $H$ is presented in figure~\ref{f:phase}. We will argue in this section that only the region of $\dot\phi$ to the right of the fixed point $\dot\phi_*$, where $J$ is positive, is healthy and relevant for the discussion of DE. 

Figure~\ref{f:phase} was obtained assuming the branch choice of
\begin{align}
	H &= \mu \dot\phi X + \sqrt{2\mu^2 X^3 + \frac{1}{3}\left(\rho_\ext-X\right)} \label{e:Hlin}\\
	J & = \dot\phi(3\mu H\dot\phi -1) \,. \label{e:Jattr2}
\end{align}
Choosing the other branch is equivalent to mapping $\dot\phi \rightarrow -\dot\phi$, $H \rightarrow -H$ and $J\rightarrow -J$. Therefore, for every expanding cosmology on this phase plot, there exists an equivalent contracting one with the opposite velocity of $\phi$.

The evolution of the background will proceed toward one of the two attractors which can be located by solving Eq.~\eqref{e:Jattr2} for $J=0$. We obtain for their positions
\begin{align}\label{Xattr}
  3\dot\phi_* H\mu =1 \qquad X_* &= (18\mu^2H^2)^{-1}\,, \\
  \text{and} \quad \dot\phi &= 0 \,.
\end{align}
On this attractor the Friedmann equation can be written in terms of the external energy density only
\begin{equation}
H^{2}=\frac{1}{6}\left(\rho_{\text{ext}}+\sqrt{\rho_{\text{ext}}^{2}+\frac{2}{3}\mu^{-2}}\right)\,.
\end{equation}
An observer who is not aware of the existence of $\phi$ and the attractor would be highly confused by this modification of the cosmological dynamics.  

Since $J$ must approach zero in an expanding universe, the basins of attraction will be delimited by the extrema of $J(\dot\phi)$. The system cannot penetrate these separatrix lines in the phase space since pressure is singular there: as already discussed in section~\S\ref{s:approach}, the normalisation of the kinetic term for perturbations, $D$, is proportional to $J_{\dot\phi}/H$ and vanishes on these extrema, causing both $\ddot{\phi}$ and the sound speed to diverge. Also, this implies that the basin of attraction around $\dot\phi=0$, since $J$ has a negative gradient there, has ghosty perturbations.

For all the values of $\dot\phi<(6\mu H)^{-1}$ (to the left of the blue line in Fig.~\ref{f:phase}), the energy density of the imperfect scalar is negative and we will not consider these phase-space regions further, since this cannot give the desired dark-energy phenomenology. The remainder of the negative-$J$ region does have positive energy density but the energy that can be stored in the scalar in the far past can only be very small and to all intents and purposes the dynamics would have been indistinguishable from the attractor for all of the observable history of the universe. 

Hence, our discussion is going to concentrate on the region $\dot\phi \geq \dot\phi_*$ where both $J$ and the scalar energy density, $\epsilon$, are positive, the perturbations are not ghostly and the background evolution in an expanding universe occurs toward $\dot\phi_*$ from above. We will show that in this region there are no gradient instabilities and the scalar field's evolution provides a viable and interesting model for dark energy.

\EPSFIGURE{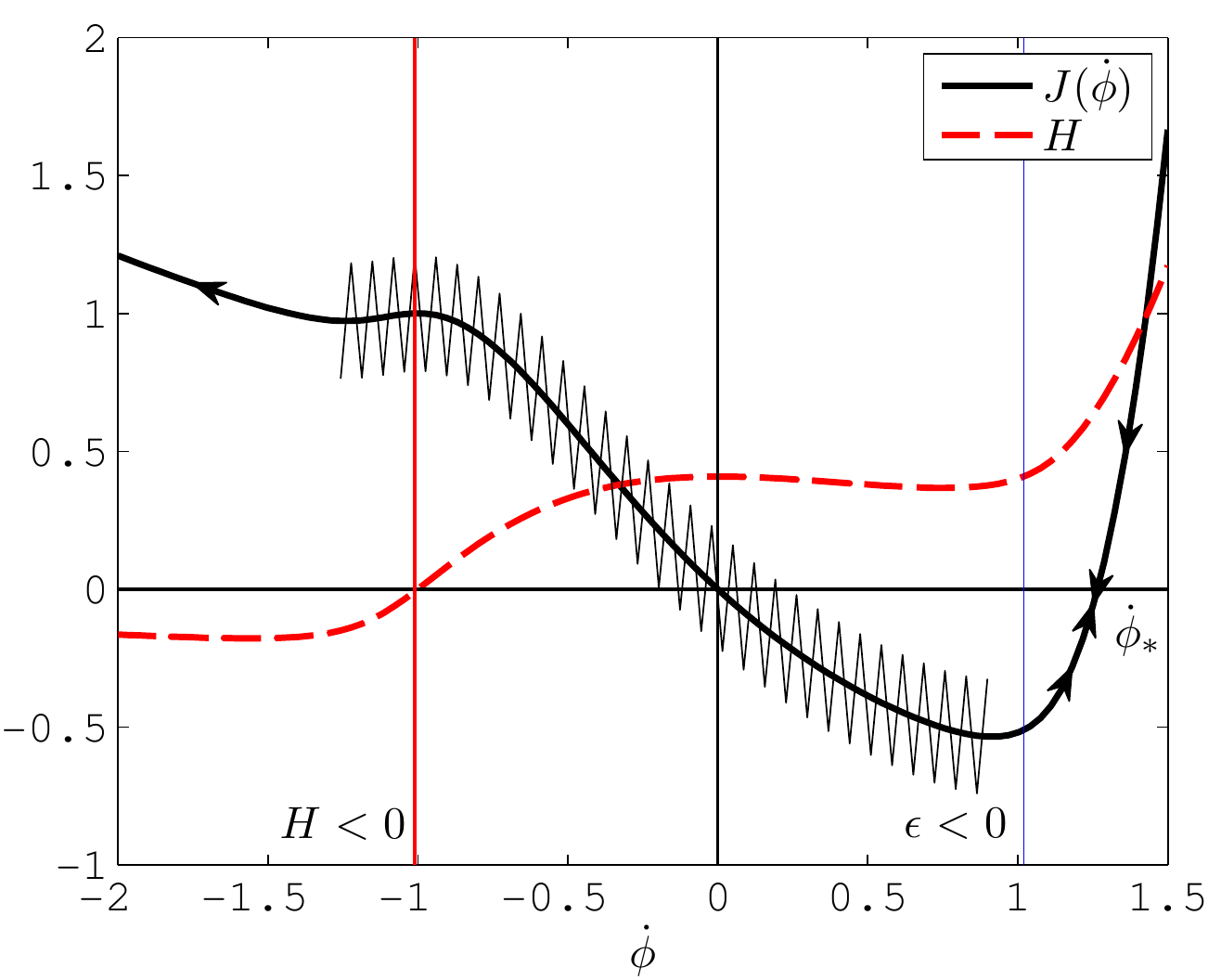}{\label{f:phase} Plot of the linear model's shift-charge density $J$ (black solid line) and the Hubble parameter $H$ (red dashed line) as a function of $\dot\phi$, for constant $\rho_\ext$. $H$ is negative to the right of the red vertical line while the energy density in the scalar is negative in the whole region to the left of the blue vertical line. The jagged line marks the region where the perturbations are ghosty. Since $D\propto J_{\dot\phi}H^{-1}$, there are only two ghost-free regions: that around $\dot\phi_*$ with $J_{\dot\phi}>0$ and $H>0$ and that to the right of the diagram with $J_{\dot\phi}<0$ and $H<0$.}

\subsection{Attractor Behaviour}
\label{s:attr}
In this section, we will assume that the initial value of $J$ was small enough such that the behaviour of the scalar at times relevant for our observations of the cosmology is effectively that of the attractor. As discussed in section~\S\ref{s:linphase}, we will focus solely on the healthy attractor at $\dot\phi_*$.

On the attractor, since $J=0$, the Friedmann equation simplifies. From Eq.~\eqref{e:H2attr}, 
\begin{equation}
  3H^2 = X_* + \rho_\ext \,, \label{e:linHattr}
\end{equation}
allowing us to define the cosmological quantities
\begin{equation}
    \Omega_X \equiv \frac {X_*}{3H^2} = (54\mu^2 H^4)^{-1}\qquad \quad \Omega_\ext \equiv \frac{\rho_\ext}{3H^2}\,,
\end{equation}
representing the contributions to the total energy density of the imperfect scalar and external matter, respectively.

The Friedmann equation Eq.~\eqref{e:Hlin} implies that when the Universe empties and $\rho_\ext\rightarrow 0$, the requirement to keep the content of the square root positive bounds $X$ from below,
\begin{equation}
  X_*^2 > \frac{1}{6\mu^2} \,.
\end{equation}
this puts a lower limit on $H$, i.e
\begin{equation}
  H^2 > \frac{1}{3\sqrt{6}\mu} \,.
\end{equation}
The somewhat counterintuitive result is that a small parameter $\mu$, i.e.\ if the  $G(X)\Box\phi$ operator is suppressed by a large mass scale, results in a universe with a large effective cosmological constant and therefore is excluded by observation. The mass scale suppressing the new term must in fact be \emph{small}\footnote{It is of course quite possible to have a very large mass scale suppressing $G\Box\phi$ provided that the final vacuum is the trival Minkowski one and dark energy is a result of a cosmological constant}. The only way of evading this constraint is by introducing an external cosmological constant which will relax the bound, also from below. If there is no external cosmological constant, and the current acceleration of the universe is being driven by an attractor such as the one discussed then the energy scale suppressing the $G\Box\phi$ term is
\begin{equation}
    \mu^{-1/3} \sim (H_0^2\Mpl)^{1/3} \sim 10^{-13} \text{eV}\,.
\end{equation}
Despite being small this mass scale has the benefit of being technically natural: the radiative corrections to this term should be small since it is an irrelevant operator, as opposed to quintessence where the scalar mass suffers from quadratic divergences.

We can explicitly confirm that the perturbations on the attractor have a positive kinetic term. Evaluating Eq.~\eqref{e:D} for the linear model at $\dot\phi_*$ we obtain
\begin{equation}
  D_* = 1 + \frac{1}{54\mu^2 H^4} = 1  + \Omega_X \,. \label{e:Dattr}
\end{equation}

Now we turn to the study of the equation of state of the imperfect scalar.
The assumption that $X$ moves on the attractor implies that
\begin{equation}
    \dot{X_*} = ( (6\mu^2H^2)^{-1}\dot{)}= -2X_*\frac{\dot{H}}{H} \,.
\end{equation}
The presence of the coupling to the Ricci tensor in the scalar equation of motion Eq.~\eqref{e:phi} resulting from gravity braiding implies that the scalar is sensitive to the presence of external matter. This is manifested by the dependence of the position of the attractor on the evolution of the Hubble parameter. Differentiating Eq.~\eqref{e:linHattr} with respect to time leads to
\begin{equation}
  \frac{\dot{H}}{H^2} = -\frac{\Omega_\ext}{2}(1+w_\ext) - \Omega_X\frac{\dot{H}}{H^2} \,.
\end{equation}
Solving for $\dot{H}/H^2$, we can write down the equation-of-state parameter for the scalar:
\begin{equation}
    1 + w^*_X = -(1+w_\ext)\frac{1-\Omega_X}{1+\Omega_X}
\end{equation}
which matches the result we could have obtained from Eq.~\eqref{w*}. In particular, we find the following limits
\begin{align}
    1+ w^*_X &\approx -(1+w_\ext) & \Omega_X \ll 1 \label{e:antitrack}\\
    1+ w^*_X &\approx -(1+w_\ext)\frac{\Omega_\ext}{2} \approx 0 & \Omega_\ext \ll 1
\end{align}
As we discussed in section \S\ref{s:attractor}, while the imperfect scalar is on its attractor, it behaves as a phantom and its energy density grows with time. In particular, while subdominant, it \emph{anti-tracks} the equation of state of the dominant energy source, by which we mean that its equation of state is related to that of the external matter by Eq.~\eqref{e:antitrack}. As the dark energy begins to dominate the energy budget, it approaches the de-Sitter equation of state from below, where it remains as a pure de-Sitter condensate.

It is important to stress that while this behaviour is phantom, the perturbations to the fluid with \emph{kinetic braiding} are not ghosty, by virtue of Eq.~\eqref{e:Dattr}. In addition, the fluid is free of gradient instabilities: the speed of sound squared is positive. Using the results of section \S\ref{s:stability}, we obtain for the sound speed on the attractor
\begin{align}
    c_{\ssl*}^2 &= \frac{1-\Omega_X}{3(1+\Omega_X)} + (1+w_\ext)\frac{1-\Omega_X}{(1+\Omega_X)^2} \,,
\end{align}
with the following limits in the early and late universe
\begin{align}
	c_{\ssl*}^2 &= \frac{4}{3} + w_\ext &\Omega_X \ll 1\,,\\
	c_{\ssl*}^2 &= \left(\frac{5}{3}+w_\ext\right)\frac{\Omega_\ext}{4}  &\Omega_\ext \ll 1 \,.
\end{align}
The speed of sound at the attractor is always positive and is determined by the external energy density. 

The history of evolution of the dark energy is presented in figure~\ref{f:attrevol} while the hydrodynamical properties of the attractor of the fluid of the imperfect scalar are summarised in table~\ref{t:hydroattr}.

\EPSFIGURE{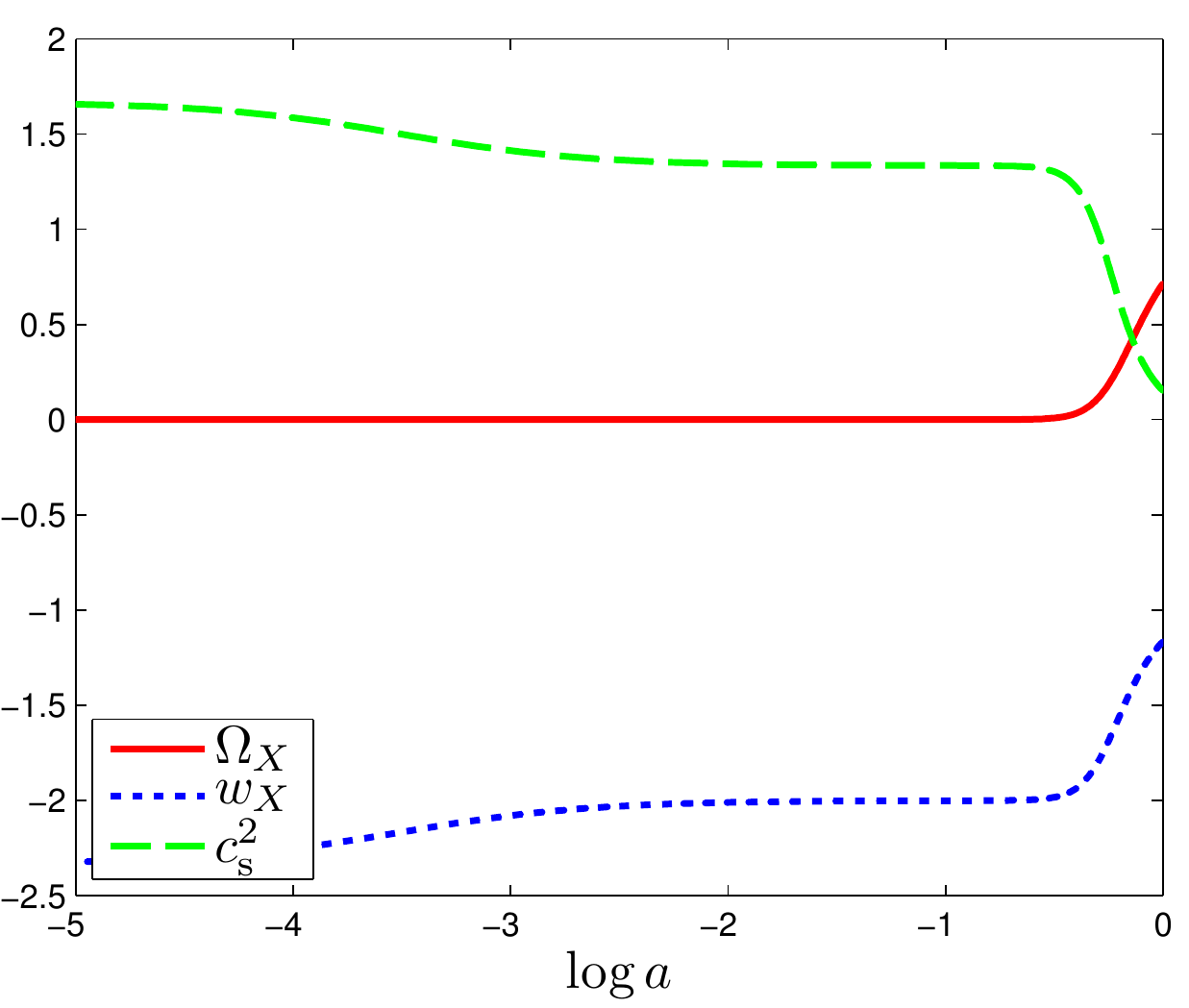}{Evolution of dark energy properties in a universe containing dust, radiation and the linear imperfect scalar evolving on its attractor throughout the presented period. During matter domination $w_X=-2$, while $w_X=-7/3$ during radiation domination. The sound speed is superluminal when the scalar energy density is subdominant, becoming subluminal when $\Omega_X \approx 0.1$ and $w_X \approx -1.4$. It's interesting to note that a model  \cite{DeFelice:2010pv} based on higher-order covariant galileon terms exhibits the same behaviour of the its equation of state as Imperfect Dark Energy. \label{f:attrevol}}

\TABULAR{rrrrrr}{
\hline\hline
\multicolumn{6}{l}{$w_X$}\\
 && $\Omega_X \approx 0$& && $\Omega_X \approx 1$\\
\hline
\\[-1ex]
 MD: && $-2$ &&& $-1 - \Omega_\ext/2$ \\[1ex]
 RD: && $-7/3$ &&& $-1 - \frac{2}{3}\Omega_\ext$ \\[1ex]
 QdS: && $-1 -(1+w_\ext)$ &&& $-1 - \frac{1}{2}(1+w_\ext)\Omega_\ext$\\[1ex]
\hline\\[-2ex]
\multicolumn{6}{l}{$c_\ssl^2$}\\[1ex]
\hline\\[-1ex]
 MD: && $4/3$ &&& $\frac{5}{12}\Omega_\ext$ \\[1ex]
 RD: && $5/3$ &&& $\frac{1}{2}\Omega_\ext$\\[1ex]
 QdS: && $\frac{1}{3}+(1+w_\ext)$ &&& $\frac{\Omega_\ext}{6}\left(2 + 3(1+w_\ext)\right)$\\[1ex]
\hline\hline
}
{Limiting values of the equation of state parameter and the speed of sound for the scalar evolving on its attractor. \label{t:hydroattr} MD and RD refer to the domination of the external fluid by matter or radiation, respectively, QdS is the inflationary quasi-de-Sitter phase.}

From the point of view of structure formation, the above results show how fundamentally different the gravity-brai\-ded fluid is to a cosmological constant: once the contribution of the external matter decreases significantly below $\Omega_\ext = 1$, the sound speed of the fluid will begin to decrease. The perturbations begin to propagate subluminally once $\Omega_X > 0.1$. As the fluid evolves towards its final de-Sitter state, the speed of sound decreases allowing it to collapse and cluster. This clustering of the dark-energy fluid should enhance the growth of structure significantly more than is observed in standard quintessence scenarios where the sound speed is always $c^2_\ssl =1$ (see, for example Ref.~\cite{dePutter:2010vy} for detailed discussion of the impact of non-canonical sound speeds on observations).

\subsection{Approach to Attractor}
\label{s:linJ}
It is key to consider precisely the manner in which the attractor described in the previous section is approached during the evolution of the cosmology. We will evaluate the hydrodynamical parameters for the imperfect scalar whose energy density is dominated by the shift current and show that it behaves in such a way so as to rapidly dilute during a period of inflation. However, should a significant shift-current density remain after inflation, we will show that it redshifts more slowly than radiation, achieving a local maximum in energy density at matter-radiation equality, acting as a model of early dark energy. Then the energy density in the scalar dilutes with respect to the dust reappearing once the scalar evolves to the vicinity of the attractor and attaches to the solution described in section~\S\ref{s:attr}. We shall refer to the period during which the scalar's dynamics are dominated by the shift-current density contribution as $J$-domination.

For $J\neq 0$ and $\dot\phi>\dot\phi_*$, it is useful to define a dimensionless charge parameter
\begin{equation}
	Q \equiv 6\mu HJ \,.
\end{equation}
Our whole region of interest is then spanned by $Q\geq0$. It turns out that the transition from $J$-dominated to attractor-like behaviours occurs when $Q\sim 1$. 

We can re-express all the quantities using $Q$ and in particular rewrite all the dimensionless parameters of the cosmology using just $Q$ and $\Omega_X$. Then,  large-$Q$ limits can be found which represent the leading-order behaviour of $J$-dominated scalars. For example,
\begin{align}
	\dot\phi &= \frac{1}{6\mu H}(1+\sqrt{1+2Q}) \simeq \frac{\sqrt{2Q}}{6\mu H}\,, \\
	\epsilon &= \frac{1}{36\mu^2H^2}\left(1+2Q+\sqrt{1+2Q}(1+Q)\right) \simeq \frac{\sqrt{2Q^3}}{36\mu^2H^2}\,, \label{e:epsQ}\\
	D &= \sqrt{1+2Q} + \frac{\Omega_X(1+Q+\sqrt{1+2Q})}{2\sqrt{1+2Q}} \simeq \sqrt{2Q}+ \Omega_X\sqrt{\frac{Q}{8}} \,.
\end{align}
The above expression for $D$ can be clearly seen to be always positive which verifies explicitly that the perturbations are healthy in the whole of the phase space of interest.

Such manipulation makes it possible to calculate the hydrodynamic parameters of the scalar fluid during $J$-domination . In particular
\begin{align}
	w_X &= \frac{1-w_\ext(1-\Omega_X)} {4+\Omega_X}\,, \\
	c_\ssl^2 &= \frac{20 + 12w_\ext(1-\Omega_X)-(4-\Omega_X)\Omega_X} {3(4+\Omega_X)^2}\,.
\end{align}
The various limits of the above expressions are presented in table~\ref{t:hydroJ}. Most importantly, just as in the case of the attractor, $c_\ssl^2>0$ for all the phases of evolution of the cosmology, confirming that the perturbations of this fluid do not exhibit gradient instabilities. It can be seen from the expression above that the speed of sound is subluminal when the energy in the scalar field is $J$-dominated.

Another important feature of this model is that fact that it is that the scalar is able to cross $w_X=-1$: a $J$-dominated fluid has $w_X>0$; as the attractor is approached, the scalar transitions to evolve with a phantom equation of state. Finally it approaches $w_X=-1$ from \emph{below}.

\TABULAR{rrr}{
\hline\hline
\multicolumn{3}{l}{$w_X$}\\
& $\Omega_X \approx 0$ & $\Omega_X \approx 1$\\
\hline
\\[-1ex]
MD:& $1/4$ & $\frac{1}{5} \left( 1- \frac{1}{5}\Omega_\ext \right)$ \\[1ex]
RD:& $1/6$ & $\frac{1}{5} \left(1 - \frac{2}{3}\Omega_\ext \right)$ \\[1ex]
 QdS: & $\frac{1}{2}\left(1-\frac{1}{2}(1+w_\ext)\right)$ & $\frac{1}{5}\left(1-(w_\ext-\frac{1}{5}) \Omega_\ext\right)$ \\[1ex]\hline\\[-2ex]
\multicolumn{3}{l}{$c_\ssl^2$}\\[1ex]
\hline\\[-1ex]
 MD:& $5/12$ & $\frac{23}{75}-\frac{12\Omega_\ext}{125}$ \\[1ex]
 RD:& $1/2$ & $\frac{23}{75}-\frac{56\Omega_\ext}{375}$ \\[1ex]
 QdS: & $\frac{1}{6}+\frac{1}{4}(1+w_\ext)$ & $\frac{23}{75} + \frac{4\Omega_\ext}{125}(2+5(1+w_\ext))$ \\[1ex]
\hline\hline
}
{Values of the equation-of-state parameter and the speed of sound for the off-attractor scalar dominated by the shift-current $J$. \label{t:hydroJ} MD and RD refer to the domination of the external fluid by matter or radiation, respectively, QdS is the inflationary quasi-de-Sitter phase.}

We now turn to the discussion of the approach to the attractor in the early universe. We assume that the initial conditions were such that the inflaton dominated the energy density sufficiently to be able to put the universe in a quasi-de-Sitter inflationary phase. However, we will also take the initial energy density stored in the imperfect scalar to be similar to that of the inflaton, since these seem to be the most natural initial conditions.

Given the model parameters required by the observation of dark energy today, this requires that most of the energy density initially be stored in the shift-current; the contribution of the attractor's energy density during inflation is tiny (since it's inversely proportional to $H^2$, see Eq.~\eqref{Xattr}). We will now show that a short period of inflation is sufficient to dilute this energy density to the extent that the dynamics of the imperfect scalar is effectively that of the attractor (i.e.\ $Q<1$) already significantly prior to the acceleration era.

Using the large-$Q$ expression Eq.~\eqref{e:epsQ} and dropping numerical coefficients gives
\begin{equation}
	\epsilon \sim \frac{H_0^4 Q^{3/2}}{H^2} \,.
\end{equation}
Then, since $\rho_\text{infl} \sim H^2_\text{infl}$, the Hubble parameter during inflation, we have for the initial value of $Q$
\begin{equation}
	Q_\text{init} \sim (\Omega_X^\text{init})^{2/3} \left( \frac{H_\text{infl}}{H_0} \right)^{8/3}
\end{equation}
During the quasi-de-Sitter phase of inflation, $H$ remains approximately constant, so $Q =6\mu HJ \propto a^{-3}$. Therefore, given $\Ncal$ e-folds of inflation, we have for $Q$ at reheating 
\begin{equation}
	Q_\text{RH} \sim Q_\text{init}\e^{-3\Ncal}
\end{equation}
Assuming that reheating occurs without a significant change in $H$, we can then write down the value of $Q$ at matter-radiation equality, given that during radiation domination $Q\propto a^{-5} \propto H^{5/2}$,
\begin{equation}
	Q_\text{eq} \sim (\Omega_X^\text{init})^{2/3} \left( \frac{H_\text{infl}}{H_0} \right)^{1/6}\left(\frac{H_\text{eq}}{H_0}\right)^{5/2} \e^{-3\Ncal}
\end{equation}
Taking $\Omega_X^\text{init}\sim 1$ and $H_\text{infl} \sim 10^{16}$~GeV implies that already for $\Ncal \sim 17$ is $Q_\text{eq} \sim 1$. As a result, under the usual $\Ncal \sim 60$ e-folds of inflation, the scalar approaches the attractor significantly before even BBN.

It would seem natural that under any such initial conditions as those discussed above, we should just take the scalar as being on its attractor throughout the whole observable period in the evolution of the universe, and therefore for its energy density to be absolutely negligible until very recently. However, since the initial conditions are unknown, we will discuss the potential effects of non-zero $J$ in the recent universe.

After reheating, the scalar redshifts with $w_X=1/6$, i.e.\ more slowly than radiation. During matter domination this becomes $w_X = 1/4$. There is therefore a local maximum in $\Omega_X$ at matter-radiation equality, provided that the the scalar energy density is still $J$-dominated. This behaviour allows there to be a significant $\Omega_X$ at equality while evading the constraints arising from the rate of expansion during BBN. Such a contribution to early dark energy (``EDE'') could potentially be measurable: it would affect the size of the acoustic horizon for both the cosmic microwave background and baryon acoustic oscillations, and alter the rate of growth of structure at redshifts close to matter-radiation equality and substantially change predictions (see for example \cite{Doran:2001rw}). A typical evolution of the parameters for the imperfect scalar under such conditions is presented in figure~\ref{f:Jdom}.

\EPSFIGURE{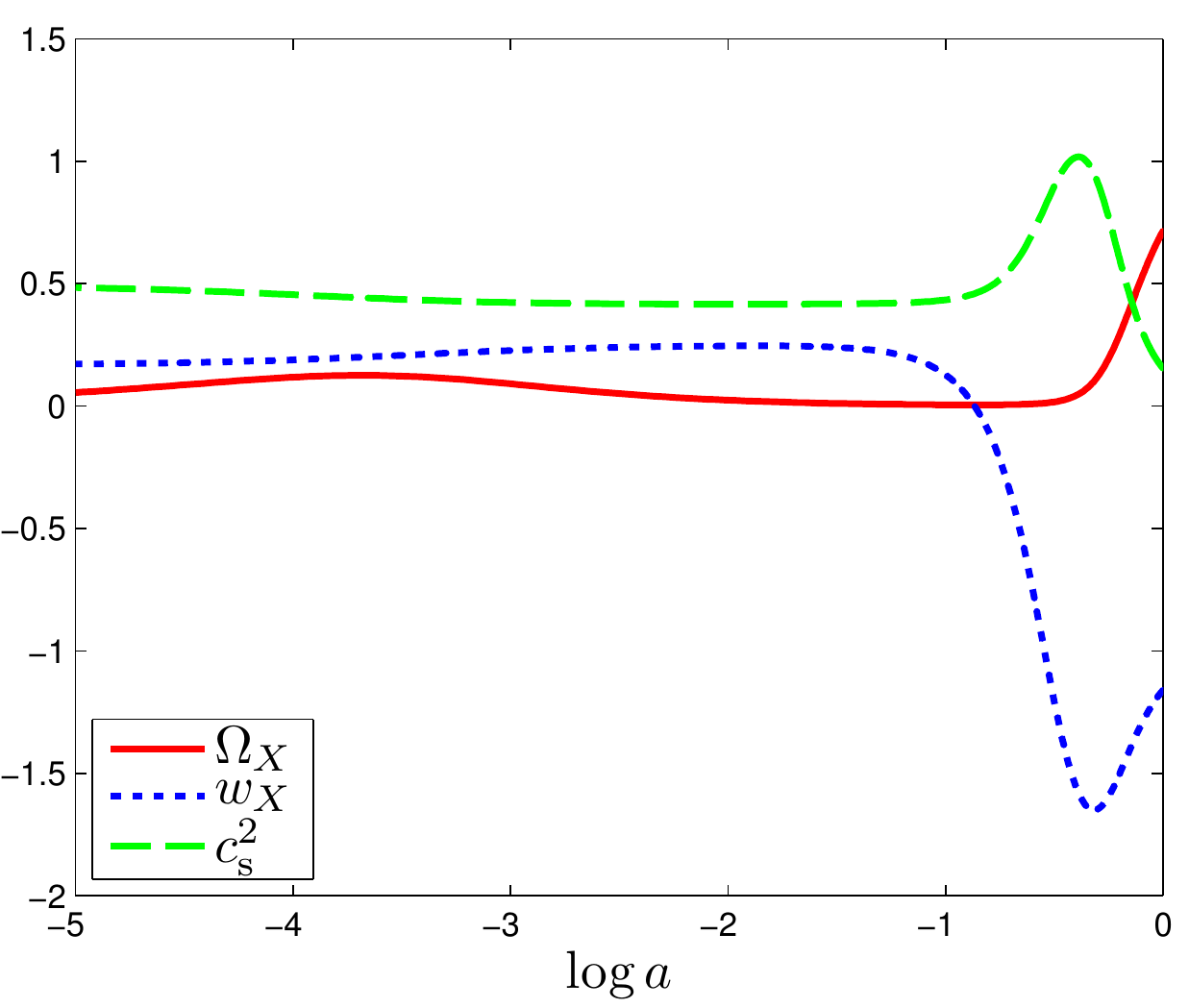}{Evolution of dark energy properties in a universe containing dust, radiation and the linear imperfect scalar. The energy density in the scalar is $J$-dominated until a transition during the matter domination epoch. This allows the scalar to increase its contribution to the total energy budget throughout radiation domination ($w_X=1/6$) and provide an early dark energy peaked at matter-radiation equality, from whence it begins to decline with $w_X=1/4$. The transition to the attractor behaviour is rapid (the time at which the transition occurs depends on the initial value of $J$). The equation of state crosses $w_X=-1$ and the scalar energy density begins to grow. The final stages of evolution are on the attractor and are effectively identical to those presented in figure~\ref{f:attrevol}. The speed of sound remains positive throughout. \label{f:Jdom}}

\subsection{Observationally Viable Parameter Space}

The linear model proposed in this section has two parameters: the value of $\mu$ and the amount of shift charge remaining today. The model can then be reparameterised in terms of the dimensionless $\Omega_{X0}$ and $Q_0$, the values of those parameters today.

A large $Q_0$ is clearly excluded, since that would give $w_X>0$ and therefore dark energy would dominate the energy density in the past. Realistic cosmologies have dark energy dominating the energy density today with small values of $Q_0$, such that the evolution is on the attractor in the recent past.

Beyond the redshift of a few, the evolution is allowed to be off-attractor, provided that the energy density $\Omega_X$ at equality is not excessive ($\Omega_X^\text{eq} \lesssim 0.1$, say\footnote{This is of the order of magnitude of the constraints given in \cite{Doran:2007ep,Xia:2009ys,dePutter:2009kn,Hollenstein:2009ph}, but clearly the actual constraints would need to be recalculated in a full analysis.}). This limits the sort of cosmologies that are permitted to be ones described by a narrow range of parameters $w_{X0}$ and $w_{a0}$. The model discussed here is not a constant $w_a$ model. However, for a convenient translation of the more frequently used parameter spaces, we include a plot of approximations to the above parameters for viable cosmologies in figures~\ref{f:OmX0}~and~\ref{f:OmXeq}.

\EPSFIGURE{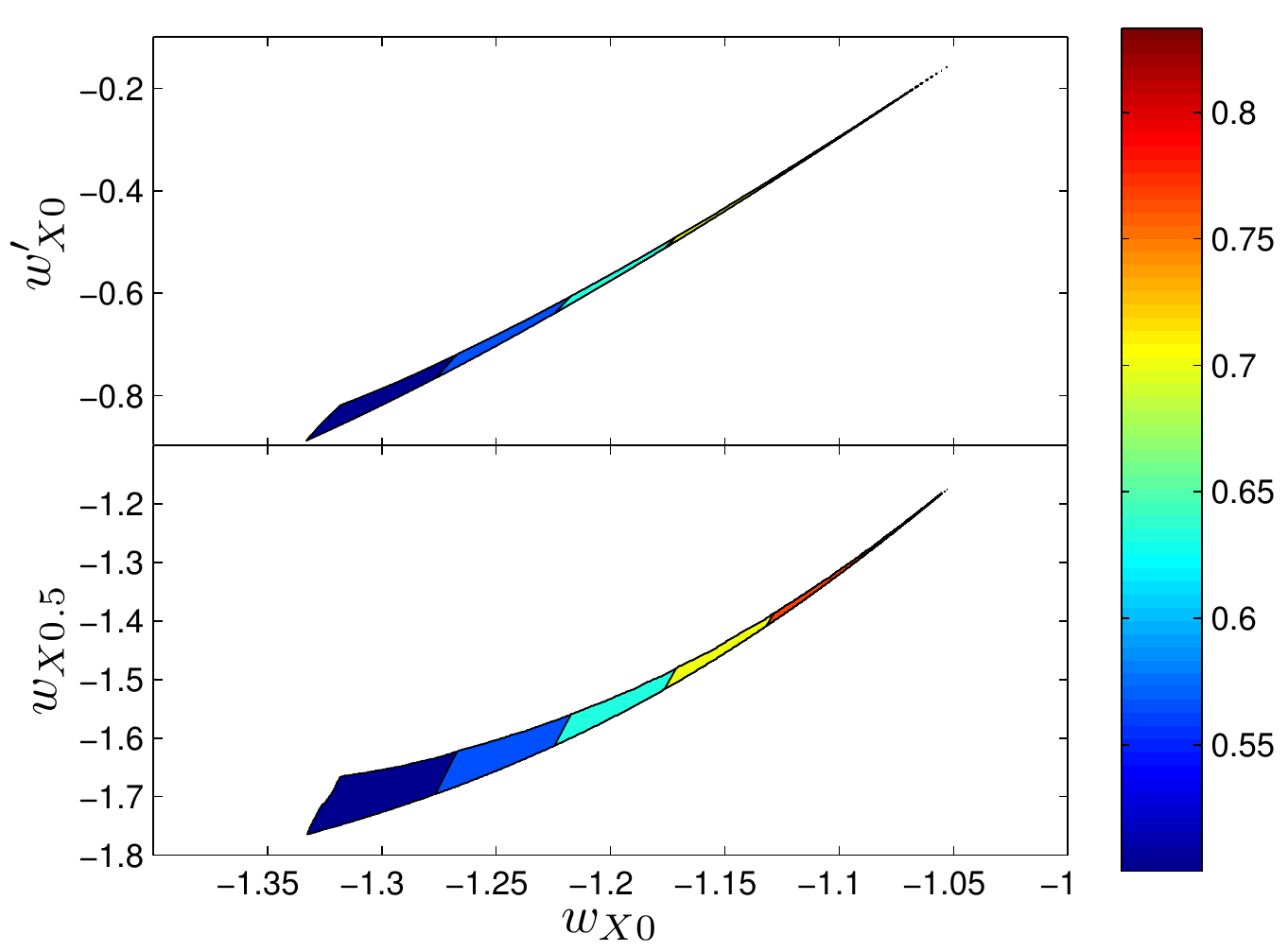}{Plot of range equation-of-state parameters allowed by requiring that the cosmological model be viable: Models are chosen by requiring that $0.1<\Omega_\ms<0.5$ and $\Omega_{X\text{eq}}<0.1$. The shading contours correspond to the energy density of dark energy today $\Omega_{X0}$. Two parameterisations of dark energy behaviour are shown: $w_X$ and $w_X'$ evaluated today, which is a reasonable approximation in the region of redshift to which supernovae are sensitive, and $w_X$ evaluated today and at $z=1/2$, which is an alternative parameterisaton \cite{Wang:2008zh}. The requirement that the energy density in the imperfect scalar at matter-radiation equality be small, $\Omega_X^\text{eq}<0.1$ forces the value of the shift charge to be small today $Q_0 < 10^{-2}$. This means that in the most recent history, the evolution has effectively been on attractor or very close to it and the permitted value of $w_X$ is very restricted and determined to all intents and purposes by $\Omega_X^0$. \label{f:OmX0}}

\EPSFIGURE{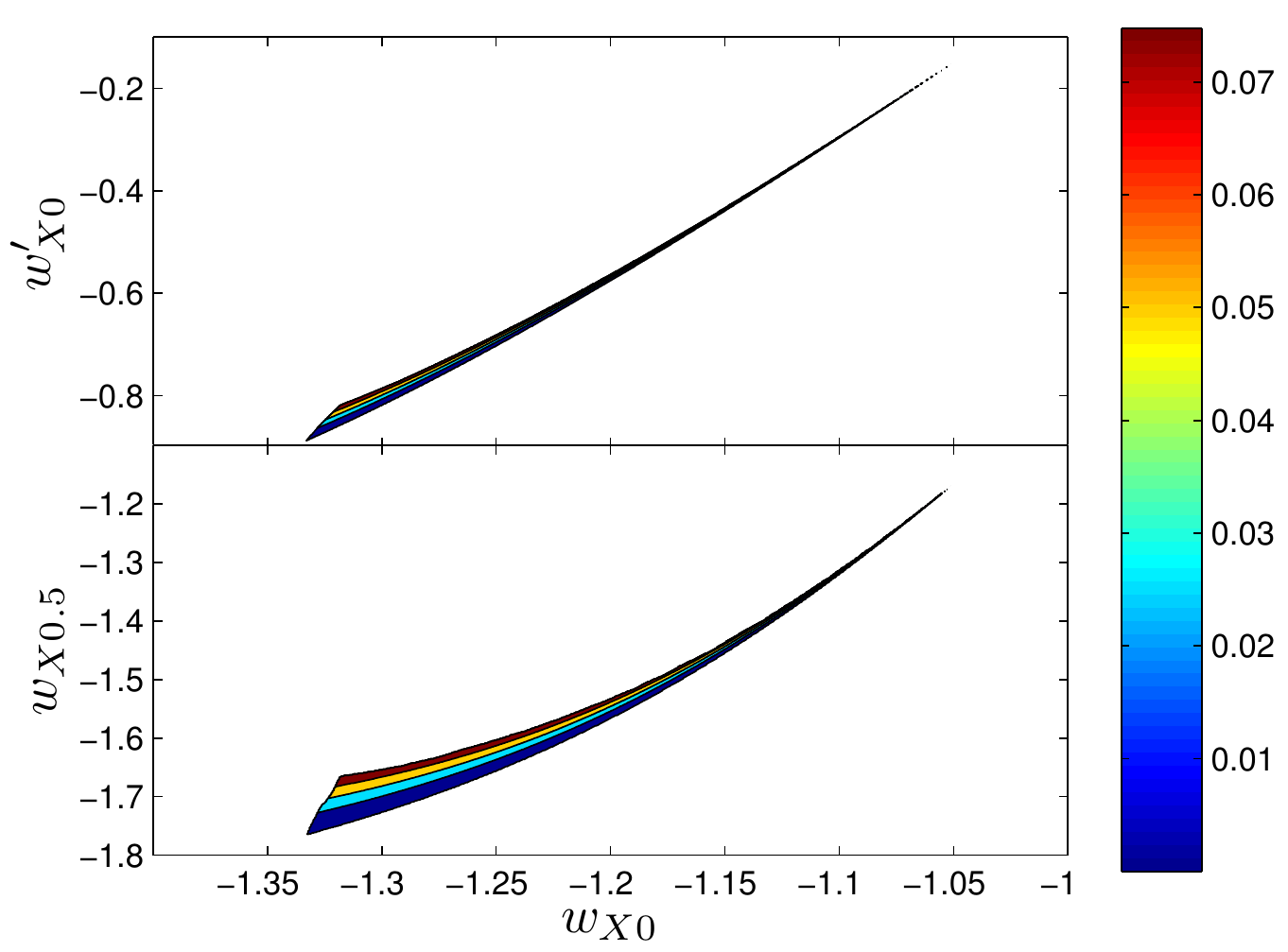}{Plot as in figure~\ref{f:OmX0}, but with shading representing the contribution of the imperfect scalar to energy density at matter-radiation equality. We choose to cut the parameters such that the contribution to this early dark energy at that time is no larger than 10\%. It can clearly be seen that values of $w_X$ closer to $-1$ are obtained when the shift charge is larger, but this leads to more early dark energy, eventually disagreeing with current constraints.\label{f:OmXeq}}

\paragraph{Note added:}
Shortly after the publication of the pre-print of this paper on the arXiv, Ref.~\cite{Kobayashi:2010cm} appeared which discusses the model \eqref{L} in the context of inflation. In this reference, the authors present the quadratic action for perturbations and find the conditions under which the perturbations remain healthy. These results agree with ours.

\acknowledgments

It is a pleasure to thank Andrei Barvinski, Gia Dvali, Gregory Gabadadze, Dmitry Gorbunov, Andrei Gruzinov, Dragan Huterer, Maxim Libanov, Eric Linder, Bob McElrath, Shinji Mukohyama, Viatcheslav Mukhanov, Carlos Pe\~{n}a Garay, Michele Redi, Jonathan Roberts, Valery Rubakov, Sergey Sibiryakov, Licia Verde, Scott Watson for very useful discussions and criticisms. The work of I.~S. and A.~V. was supported by the James Arthur Fellowship. I.~S. and A.~V.  would like to thank the CERN theory devision for their hospitality during the preparation of this manuscript. This visit of A.~V. to CERN was supported through a grant of the David and Lucile Packard Foundation.  A.~V. is also thankful to the organisers of the workshop \emph{Gravity and Cosmology 2010} at the Yukawa Institute for Theoretical Physics, Kyoto University for their hospitality and financial support during the final stages of this project. 

\appendix

\section{Action for Cosmological Perturbations}
\label{sec:apppert}

In this section, we derive the quadratic action for perturbations of the full scalar-tensor system around an arbitrary homogeneous and isotropic background. We follow the method introduced in \cite{Maldacena:2002vr}.

Let us start by writing the action Eq.~\eqref{e:action} in a slightly more convenient form. Integrating by parts twice the $G\,B $ term, one obtains
\beq\label{action2}
S={1\over2}\int \ds^4x \sqrt{-g}\lp\{ R+2K+2\theta G\sqrt{2X} + G\frac {\nabla_\mu X\nabla^\mu\phi}{2X}\rp\}\,,
\eeq
where as before $\theta\equiv \nabla_\mu u^\mu$ is the trace of the extrinsic curvature of the slicing $\phi=\text{const}$ with the normal  $u_\mu=\nabla_\mu\phi/\sqrt{2X}$.

The computation of the quadratic action is most convenient in ADM variables, where the metric is written as
\beq\label{adm}
\ds s^2=N^2 \ds t^2 - h_{ij}(N^i \ds t + \ds x^i)(N^j \ds t + \ds x^j)
\eeq
where $N$ is the lapse function, $N^i$ is the shift vector and $h_{ij}$ is the induced metric on the slices of constant time $t$. The extrinsic curvature of these slices is $E_{ij}/N$ with
$$
E_{ij}={1\over2} \dot h_{ij} - D_{(i} N_{j)}
$$
where $D_i$ is the covariant derivative associated with $h_{ij}$ and the indices are raised and lowered with $h_{ij}$.

The usefulness of these variables is that in the `unitary' gauge, defined as $\phi=\phi_0(t)$ (that is $\delta\phi=0$), the slicing defined by $t$ and $\phi$ coincide. Consequently, the $i,j$ components of the spatial projector $\proj_{\mu\nu}$ and of the extrinsic curvature $K_{\mu\nu}\equiv \proj_{(\mu}^\rho \proj_{\nu)}^\sigma \nabla_\rho u_\sigma$ coincide with $h_{ij}$ and $E_{ij}/N$ respectively. Furthermore, in this gauge, the full unperturbed form of $X$ is simply
$$
X= \dot\phi_0^2/(2N^2)~,
$$
which depends only on $\delta N$.

The background solution is given by $h_{ij} = a^2(t) \delta_{ij}$, $N=1$, $N^i=0$ and $\phi=\phi_0(t)$.
It is also convenient to decompose the spatial metric as
$$
h_{ij}=e^{2\sigma(t) + 2 \zeta(t,x)} \;\hat h_{ij}
$$
where $\sigma=\log a(t)$ and $\det(\hat h_{ij})=1$. The variable $\zeta$ is the curvature perturbation.

To fix the remaining gauge freedom it is useful to impose
$$
\partial^i \hat h_{ij}=0~,
$$
so that $\hat h_{ij}$ is a `transverse-traceless' variable, describing the tensor modes.

The action in ADM variables is
\begin{align}\label{actionADM}
S=\frac{1}{2}\int \ds^3x\ds t \sqrt{h} \,&\lp\{ N(R_3+2K(X)) + \frac{1}{N} \lp(E_{ij}^2-E^2\rp)+ \right.\\
 &\left.+2G\sqrt{2X}E + \frac{G}{NX}\left( \dot{X}\dot\phi + N^i\partial_i X \dot\phi \right)  \rp\}\notag\,,
\end{align}
where $R_3$ is the Ricci scalar for $h_{ij}$, we have dropped terms containing $\partial_i \phi$ which is always vanishing in the chosen gauge and we remind that $X=X(N)$.

The strategy devised in Ref.~\cite{Maldacena:2002vr} to obtain the quadratic action was to first compute $\delta N$ and $\delta N^i=N^i$ at first order in terms of $\zeta$ and $\hat h_{ij}$ using the Hamiltonian and momentum constraints and then to substitute them in the action and expand it to quadratic order. The expression of $\delta N$ and $N^i$ to second order is not necessary because those terms are multiplied by the constraints at zeroth order, which identically vanish.

The momentum constraint is
$$
D^i\lp( {E_{ij}-E\,h_{ij} \over N} + G\sqrt{2X} h_{ij}\rp) -\frac{G}{2NX} \partial_iX \dot\phi=0~.
$$
It is convenient to separate the shift vector into its transverse and longitudinal parts, $N_i = N_i^{T}+ \partial_i \psi$ with $D^i N_i^T=0$. Expanding the momentum constraint at linear order, its
transverse and longitudinal parts give
\beq\label{NT}
N_i^{T}=0
\eeq
and
\beq\label{deltaN}
(H-G_X X\sqrt{2X})|_0 \delta N = \dot \zeta\,,
\eeq
respectively. $\psi$ does not appear in the quadratic action since it enters the action \eqref{actionADM} through $N^i$, which only appears (up to total derivatives) in the combination $N^i$ times the linearised momentum constraint, which is set to vanish. As usual, the subscript ${}_0$ denotes the background value, e.g., $X_0=\dot\phi_0^2/2$ etc. Notice here that the relationship between the perturbations of the lapse and the curvature, Eq.~\eqref{deltaN}, is changed from the one in k-\emph{essence}. One could expect that this will impact the normalisations of inflationary perturbations in a new way.

Substituting the momentum constraint into the action, dropping terms which will only contribute at orders above second and integrating by parts we obtain the partially reduced action
\begin{equation}
  S = \frac{1}{2}\int \sqrt{h} \left[N(R_3 + 2K) + \frac{1}{N}\left(A_{ij}^2-A^2\right) -2\dot{G}\sqrt{2X} \right]\,.
\end{equation}
where $A_{ij} \equiv \dot{h}_{ij}/2$. Substituting for perturbed variables we find:
\begin{align}\label{reducedaction}
S = \frac{1}{2} \int& \ds^3x \ds t \; e^{3(\sigma+\zeta)}
\Big\{(1+\delta N) 
\Big(2 K(\phi,X(1+\delta N)) e^{-2(\sigma+\zeta)}
\Big[ 4\Delta\zeta + 2 (\partial\zeta)^2 + {(\partial\hat h)^2\over 4} \Big]
\Big)\notag\\
&+\frac{1}{1+\delta N} \Big( - 6(H+\dot\zeta)^2 + \frac{ (\dot{\hat h})^2}{4}\Big)
-2 \dot{G}(\phi,X(1+\delta N))\sqrt{2X(1+\delta N)}\Big\} \,.
\end{align}

Expanding to second order, after some manipulations and using the background equations of motion, one finds the quadratic action
\beq\label{quadraticaction}
S^{(2)}=\int \ds^3x\ds t a^3\lp\{ {A\over2} \lp[\dot\zeta^2 - {c_\ssl^2 \over a^2} (\partial_i\zeta)^2  \rp]
+ {1\over8}\lp[ (\dot{\hat h}_{ij})^2 - {1 \over a^2}(\partial_k \hat h_{ij})^2\rp] \rp\}~,
\eeq
where the normalization of the kinetic term is
\begin{align}\label{A_app}
A&={2XD\over (H-\dot\phi XG_X)^2}\\
D&=K_X+2XK_{XX}- 2G_\phi - 2XG_{X\phi} +6\dot\phi H(G_X+XG_{XX})+6X^2G_X^2
\end{align}
and the sound speed for the scalar perturbations is
\beq\label{cs_app}
c_\ssl^2=
{\dot\phi XG_X (H- \dot\phi XG_X)-(H-\dot\phi X G_X)^{{}^\centerdot} \over
XD}
\eeq
where it is understood that everything is evaluated on the background solution. The Friedmann equations then have to be used to eliminate $\dot{H}$, remembering that this derivation was performed under the assumption of the imperfect scalar's providing the total energy content of the cosmology. In this way the above reduces to
\begin{equation}\label{cs2_app}
c_\ssl^2 = \frac{K_X-2G_\phi + 2XG_{\phi X} + 2\ddot\phi(G_X+XG_{XX}) + 4\dot\phi HG_X - 2X^2G_X^2}{D} \,.
\end{equation}
The dependence of the sound speed on $\ddot\phi$ remains and the equation of motion for the imperfect scalar has to be used in order to complete the evaluation. Generically, therefore, the sound speed will depend on the dynamics of the whole system and is not just a property of the fluid.

Finally, note that the properties of the tensor modes are not affected. The sound speed is still 1, as expected because we are only modifying gravity through the trace of the extrinsic curvature, which is independent of $\hat h_{ij}$.

\bibliography{KGB}

\providecommand{\href}[2]{#2}\begingroup\raggedright\begin{thebibliography}{10%
0}

\bibitem{Dvali:2000hr}
G.~R. Dvali, G.~Gabadadze, and M.~Porrati, ``{4D gravity on a brane in 5D
  Minkowski space},''
  \href{http://dx.doi.org/10.1016/S0370-2693(00)00669-9}{{\em Phys. Lett.} {\bf
  B485} (2000)  208--214},
\href{http://arxiv.org/abs/hep-th/0005016}{{\tt arXiv:hep-th/0005016}}.

\bibitem{deRham:2007xp}
C.~de~Rham {\em et al.}, ``{Cascading gravity: Extending the
  Dvali-Gabadadze-Porrati model to higher dimension},''
  \href{http://dx.doi.org/10.1103/PhysRevLett.100.251603}{{\em Phys. Rev.
  Lett.} {\bf 100} (2008)  251603},
\href{http://arxiv.org/abs/0711.2072}{{\tt arXiv:0711.2072 [hep-th]}}.

\bibitem{ArkaniHamed:2003uy}
N.~Arkani-Hamed, H.-C. Cheng, M.~A. Luty, and S.~Mukohyama, ``{Ghost
  condensation and a consistent infrared modification of gravity},'' {\em JHEP}
  {\bf 05} (2004)  074,
\href{http://arxiv.org/abs/hep-th/0312099}{{\tt arXiv:hep-th/0312099}}.

\bibitem{Rubakov:2004eb}
V.~A. Rubakov, ``{Lorentz-violating graviton masses: Getting around ghosts, low
  strong coupling scale and VDVZ discontinuity},''
\href{http://arxiv.org/abs/hep-th/0407104}{{\tt arXiv:hep-th/0407104}}.

\bibitem{Dubovsky:2004sg}
S.~L. Dubovsky, ``{Phases of massive gravity},''
  \href{http://dx.doi.org/10.1088/1126-6708/2004/10/076}{{\em JHEP} {\bf 10}
  (2004)  076},
\href{http://arxiv.org/abs/hep-th/0409124}{{\tt arXiv:hep-th/0409124}}.

\bibitem{Rubakov:2008nh}
V.~A. Rubakov and P.~G. Tinyakov, ``{Infrared-modified gravities and massive
  gravitons},'' \href{http://dx.doi.org/10.1070/PU2008v051n08ABEH006600}{{\em
  Phys. Usp.} {\bf 51} (2008)  759--792},
\href{http://arxiv.org/abs/0802.4379}{{\tt arXiv:0802.4379 [hep-th]}}.

\bibitem{lpr}
M.~A. Luty, M.~Porrati, and R.~Rattazzi, ``{Strong interactions and stability
  in the DGP model},'' {\em JHEP} {\bf 09} (2003)  029,
\href{http://arxiv.org/abs/hep-th/0303116}{{\tt arXiv:hep-th/0303116}}.

\bibitem{Nicolis:2008in}
A.~Nicolis, R.~Rattazzi, and E.~Trincherini, ``{The galileon as a local
  modification of gravity},''
  \href{http://dx.doi.org/10.1103/PhysRevD.79.064036}{{\em Phys. Rev.} {\bf
  D79} (2009)  064036},
\href{http://arxiv.org/abs/0811.2197}{{\tt arXiv:0811.2197 [hep-th]}}.

\bibitem{covGal}
C.~Deffayet, G.~Esposito-Farese, and A.~Vikman, ``{Covariant Galileon},''
  \href{http://dx.doi.org/10.1103/PhysRevD.79.084003}{{\em Phys. Rev.} {\bf
  D79} (2009)  084003},
\href{http://arxiv.org/abs/0901.1314}{{\tt arXiv:0901.1314 [hep-th]}}.

\bibitem{ArkaniHamed:2002sp}
N.~Arkani-Hamed, H.~Georgi, and M.~D. Schwartz, ``{Effective field theory for
  massive gravitons and gravity in theory space},'' {\em Ann. Phys.} {\bf 305}
  (2003)  96--118,
\href{http://arxiv.org/abs/hep-th/0210184}{{\tt arXiv:hep-th/0210184}}.

\bibitem{Deffayet:2005ys}
C.~Deffayet and J.-W. Rombouts, ``{Ghosts, strong coupling and accidental
  symmetries in massive gravity},''
  \href{http://dx.doi.org/10.1103/PhysRevD.72.044003}{{\em Phys. Rev.} {\bf
  D72} (2005)  044003},
\href{http://arxiv.org/abs/gr-qc/0505134}{{\tt arXiv:gr-qc/0505134}}.

\bibitem{Creminelli:2005qk}
P.~Creminelli, A.~Nicolis, M.~Papucci, and E.~Trincherini, ``{Ghosts in massive
  gravity},'' {\em JHEP} {\bf 09} (2005)  003,
\href{http://arxiv.org/abs/hep-th/0505147}{{\tt arXiv:hep-th/0505147}}.

\bibitem{nr}
A.~Nicolis and R.~Rattazzi, ``{Classical and quantum consistency of the DGP
  model},'' {\em JHEP} {\bf 06} (2004)  059,
\href{http://arxiv.org/abs/hep-th/0404159}{{\tt arXiv:hep-th/0404159}}.

\bibitem{Babichev:2009ee}
E.~Babichev, C.~Deffayet, and R.~Ziour, ``{k-Mouflage gravity},''
  \href{http://dx.doi.org/10.1142/S0218271809016107}{{\em Int. J. Mod. Phys.}
  {\bf D18} (2009)  2147--2154},
\href{http://arxiv.org/abs/0905.2943}{{\tt arXiv:0905.2943 [hep-th]}}.

\bibitem{Babichev:2009us}
E.~Babichev, C.~Deffayet, and R.~Ziour, ``{The Vainshtein mechanism in the
  Decoupling Limit of massive gravity},''
  \href{http://dx.doi.org/10.1088/1126-6708/2009/05/098}{{\em JHEP} {\bf 05}
  (2009)  098},
\href{http://arxiv.org/abs/0901.0393}{{\tt arXiv:0901.0393 [hep-th]}}.

\bibitem{Babichev:2010jd}
E.~Babichev, C.~Deffayet, and R.~Ziour, ``{The recovery of General Relativity
  in massive gravity via the Vainshtein mechanism},''
\href{http://arxiv.org/abs/1007.4506}{{\tt arXiv:1007.4506 [gr-qc]}}.

\bibitem{Fairlie:1991qe}
D.~B. Fairlie, J.~Govaerts, and A.~Morozov, ``{Universal Field Equations with
  Covariant Solutions D.B. Fairlie},''
  \href{http://dx.doi.org/10.1016/0550-3213(92)90455-K}{{\em Nucl. Phys.} {\bf
  B373} (1992)  214--232},
\href{http://arxiv.org/abs/hep-th/9110022}{{\tt arXiv:hep-th/9110022}}.

\bibitem{Fairlie:1992nb}
D.~B. Fairlie and J.~Govaerts, ``{Euler hierarchies and universal equations},''
  \href{http://dx.doi.org/10.1063/1.529904}{{\em J. Math. Phys.} {\bf 33}
  (1992)  3543--3566},
\href{http://arxiv.org/abs/hep-th/9204074}{{\tt arXiv:hep-th/9204074}}.

\bibitem{Deffayet:2010zh}
C.~Deffayet, S.~Deser, and G.~Esposito-Farese, ``{Arbitrary p-form
  Galileons},''
\href{http://arxiv.org/abs/1007.5278}{{\tt arXiv:1007.5278 [gr-qc]}}.

\bibitem{Deffayet:2009mn}
C.~Deffayet, S.~Deser, and G.~Esposito-Farese, ``{Generalized Galileons: All
  scalar models whose curved background extensions maintain second-order field
  equations and stress-tensors},''
  \href{http://dx.doi.org/10.1103/PhysRevD.80.064015}{{\em Phys. Rev.} {\bf
  D80} (2009)  064015},
\href{http://arxiv.org/abs/0906.1967}{{\tt arXiv:0906.1967 [gr-qc]}}.

\bibitem{deRham:2010eu}
C.~de~Rham and A.~J. Tolley, ``{DBI and the Galileon reunited},''
  \href{http://dx.doi.org/10.1088/1475-7516/2010/05/015}{{\em JCAP} {\bf 1005}
  (2010)  015},
\href{http://arxiv.org/abs/1003.5917}{{\tt arXiv:1003.5917 [hep-th]}}.

\bibitem{deRham:2010gu}
C.~de~Rham and G.~Gabadadze, ``{Selftuned Massive Spin-2},''
  \href{http://dx.doi.org/10.1016/j.physletb.2010.08.043}{{\em Phys. Lett.}
  {\bf B693} (2010)  334--338},
\href{http://arxiv.org/abs/1006.4367}{{\tt arXiv:1006.4367 [hep-th]}}.

\bibitem{deRham:2010ik}
C.~de~Rham and G.~Gabadadze, ``{Generalization of the Fierz-Pauli Action},''
  \href{http://dx.doi.org/10.1103/PhysRevD.82.044020}{{\em Phys. Rev.} {\bf
  D82} (2010)  044020},
\href{http://arxiv.org/abs/1007.0443}{{\tt arXiv:1007.0443 [hep-th]}}.

\bibitem{ArmendarizPicon:1999rj}
C.~Armendariz-Picon, T.~Damour, and V.~F. Mukhanov, ``{k-Inflation},''
  \href{http://dx.doi.org/10.1016/S0370-2693(99)00603-6}{{\em Phys. Lett.} {\bf
  B458} (1999)  209--218},
\href{http://arxiv.org/abs/hep-th/9904075}{{\tt arXiv:hep-th/9904075}}.

\bibitem{ArmendarizPicon:2000dh}
C.~Armendariz-Picon, V.~F. Mukhanov, and P.~J. Steinhardt, ``{A dynamical
  solution to the problem of a small cosmological constant and late-time cosmic
  acceleration},'' \href{http://dx.doi.org/10.1103/PhysRevLett.85.4438}{{\em
  Phys. Rev. Lett.} {\bf 85} (2000)  4438--4441},
\href{http://arxiv.org/abs/astro-ph/0004134}{{\tt arXiv:astro-ph/0004134}}.

\bibitem{ArmendarizPicon:2000ah}
C.~Armendariz-Picon, V.~F. Mukhanov, and P.~J. Steinhardt, ``{Essentials of
  k-essence},'' \href{http://dx.doi.org/10.1103/PhysRevD.63.103510}{{\em Phys.
  Rev.} {\bf D63} (2001)  103510},
\href{http://arxiv.org/abs/astro-ph/0006373}{{\tt arXiv:astro-ph/0006373}}.

\bibitem{Imperfect}
C.~Deffayet, O.~Pujolas, I.~Sawicki, and A.~{Vikman,} {\em in preparation}  .

\bibitem{Komatsu:2010fb}
E.~Komatsu {\em et al.}, ``{Seven-Year Wilkinson Microwave Anisotropy Probe
  (WMAP) Observations: Cosmological Interpretation},''
\href{http://arxiv.org/abs/1001.4538}{{\tt arXiv:1001.4538 [astro-ph.CO]}}.

\bibitem{Chow:2009fm}
N.~Chow and J.~Khoury, ``{Galileon Cosmology},''
  \href{http://dx.doi.org/10.1103/PhysRevD.80.024037}{{\em Phys. Rev.} {\bf
  D80} (2009)  024037},
\href{http://arxiv.org/abs/0905.1325}{{\tt arXiv:0905.1325 [hep-th]}}.

\bibitem{Silva:2009km}
F.~P. Silva and K.~Koyama, ``{Self-Accelerating Universe in Galileon
  Cosmology},'' \href{http://dx.doi.org/10.1103/PhysRevD.80.121301}{{\em Phys.
  Rev.} {\bf D80} (2009)  121301},
\href{http://arxiv.org/abs/0909.4538}{{\tt arXiv:0909.4538 [astro-ph.CO]}}.

\bibitem{Kobayashi:2010wa}
T.~Kobayashi, ``{Cosmic expansion and growth histories in Galileon scalar-
  tensor models of dark energy},''
  \href{http://dx.doi.org/10.1103/PhysRevD.81.103533}{{\em Phys. Rev.} {\bf
  D81} (2010)  103533},
\href{http://arxiv.org/abs/1003.3281}{{\tt arXiv:1003.3281 [astro-ph.CO]}}.

\bibitem{DeFelice:2010jn}
A.~De~Felice and S.~Tsujikawa, ``{Generalized Brans-Dicke theories},''
  \href{http://dx.doi.org/10.1088/1475-7516/2010/07/024}{{\em JCAP} {\bf 1007}
  (2010)  024},
\href{http://arxiv.org/abs/1005.0868}{{\tt arXiv:1005.0868 [astro-ph.CO]}}.

\bibitem{DeFelice:2010gb}
A.~De~Felice, S.~Mukohyama, and S.~Tsujikawa, ``{Density perturbations in
  general modified gravitational theories},''
\href{http://arxiv.org/abs/1006.0281}{{\tt arXiv:1006.0281 [astro-ph.CO]}}.

\bibitem{Creminelli:2010ba}
P.~Creminelli, A.~Nicolis, and E.~Trincherini, ``{Galilean Genesis: an
  alternative to inflation},''
\href{http://arxiv.org/abs/1007.0027}{{\tt arXiv:1007.0027 [hep-th]}}.

\bibitem{Nicolis:2009qm}
A.~Nicolis, R.~Rattazzi, and E.~Trincherini, ``{Energy's and amplitudes'
  positivity},'' \href{http://dx.doi.org/10.1007/JHEP05(2010)095}{{\em JHEP}
  {\bf 05} (2010)  095},
\href{http://arxiv.org/abs/0912.4258}{{\tt arXiv:0912.4258 [hep-th]}}.

\bibitem{DeFelice:2010pv}
A.~De~Felice and S.~Tsujikawa, ``{Cosmology of a covariant Galileon field},''
\href{http://arxiv.org/abs/1007.2700}{{\tt arXiv:1007.2700 [astro-ph.CO]}}.

\bibitem{Gannouji:2010au}
R.~Gannouji and M.~Sami, ``{Galileon gravity and its relevance to late time
  cosmic acceleration},''
  \href{http://dx.doi.org/10.1103/PhysRevD.82.024011}{{\em Phys. Rev.} {\bf
  D82} (2010)  024011},
\href{http://arxiv.org/abs/1004.2808}{{\tt arXiv:1004.2808 [gr-qc]}}.

\bibitem{Garriga:1999vw}
J.~Garriga and V.~F. Mukhanov, ``{Perturbations in k-inflation},''
  \href{http://dx.doi.org/10.1016/S0370-2693(99)00602-4}{{\em Phys. Lett.} {\bf
  B458} (1999)  219--225},
\href{http://arxiv.org/abs/hep-th/9904176}{{\tt arXiv:hep-th/9904176}}.

\bibitem{Dyer:2009yg}
E.~Dyer and K.~Hinterbichler, ``{Boundary Terms and Junction Conditions for the
  DGP pi- Lagrangian},''
  \href{http://dx.doi.org/10.1088/1126-6708/2009/11/059}{{\em JHEP} {\bf 11}
  (2009)  059},
\href{http://arxiv.org/abs/0907.1691}{{\tt arXiv:0907.1691 [hep-th]}}.

\bibitem{Sushkov:2009hk}
S.~V. Sushkov, ``{Exact cosmological solutions with nonminimal derivative
  coupling},'' \href{http://dx.doi.org/10.1103/PhysRevD.80.103505}{{\em Phys.
  Rev.} {\bf D80} (2009)  103505},
\href{http://arxiv.org/abs/0910.0980}{{\tt arXiv:0910.0980 [gr-qc]}}.

\bibitem{Saridakis:2010mf}
E.~N. Saridakis and S.~V. Sushkov, ``{Quintessence and phantom cosmology with
  non-minimal derivative coupling},''
  \href{http://dx.doi.org/10.1103/PhysRevD.81.083510}{{\em Phys. Rev.} {\bf
  D81} (2010)  083510},
\href{http://arxiv.org/abs/1002.3478}{{\tt arXiv:1002.3478 [gr-qc]}}.

\bibitem{Gao:2010vr}
C.~Gao, ``{When scalar field is kinetically coupled to the Einstein tensor},''
  \href{http://dx.doi.org/10.1088/1475-7516/2010/06/023}{{\em JCAP} {\bf 1006}
  (2010)  023},
\href{http://arxiv.org/abs/1002.4035}{{\tt arXiv:1002.4035 [gr-qc]}}.

\bibitem{Germani:2010gm}
C.~Germani and A.~Kehagias, ``{New Model of Inflation with Non-minimal
  Derivative Coupling of Standard Model Higgs Boson to Gravity},''
  \href{http://dx.doi.org/10.1103/PhysRevLett.105.011302}{{\em Phys. Rev.
  Lett.} {\bf 105} (2010)  011302},
\href{http://arxiv.org/abs/1003.2635}{{\tt arXiv:1003.2635 [hep-ph]}}.

\bibitem{Germani:2010ux}
C.~Germani and A.~Kehagias, ``{Cosmological Perturbations in the New Higgs
  Inflation},'' \href{http://dx.doi.org/10.1088/1475-7516/2010/05/019}{{\em
  JCAP} {\bf 1005} (2010)  019},
\href{http://arxiv.org/abs/1003.4285}{{\tt arXiv:1003.4285 [astro-ph.CO]}}.

\bibitem{Mukohyama:2003nw}
S.~Mukohyama and L.~Randall, ``{A dynamical approach to the cosmological
  constant},'' \href{http://dx.doi.org/10.1103/PhysRevLett.92.211302}{{\em
  Phys. Rev. Lett.} {\bf 92} (2004)  211302},
\href{http://arxiv.org/abs/hep-th/0306108}{{\tt arXiv:hep-th/0306108}}.

\bibitem{Nojiri:2004bi}
S.~Nojiri and S.~D. Odintsov, ``{Gravity assisted dark energy dominance and
  cosmic acceleration},''
  \href{http://dx.doi.org/10.1016/j.physletb.2004.08.045}{{\em Phys. Lett.}
  {\bf B599} (2004)  137--142},
\href{http://arxiv.org/abs/astro-ph/0403622}{{\tt arXiv:astro-ph/0403622}}.

\bibitem{Babichev:2009jt}
E.~Babichev, C.~Deffayet, and R.~Ziour, ``{Recovering General Relativity from
  massive gravity},''
  \href{http://dx.doi.org/10.1103/PhysRevLett.103.201102}{{\em Phys. Rev.
  Lett.} {\bf 103} (2009)  201102},
\href{http://arxiv.org/abs/0907.4103}{{\tt arXiv:0907.4103 [gr-qc]}}.

\bibitem{Kobayashi:2009wr}
T.~Kobayashi, H.~Tashiro, and D.~Suzuki, ``{Evolution of linear cosmological
  perturbations and its observational implications in Galileon-type modified
  gravity},'' {\em Phys. Rev.} {\bf D81} (2010)  063513,
\href{http://arxiv.org/abs/0912.4641}{{\tt arXiv:0912.4641 [astro-ph.CO]}}.

\bibitem{Courant}
R.~Courant and D.~Hilbert, {\em {Methods of Mathematical Physics, Vol. 2}}.
\newblock Wiley-Interscience, 1989.

\bibitem{Babichev:2007dw}
E.~Babichev, V.~Mukhanov, and A.~Vikman, ``{k-Essence, superluminal
  propagation, causality and emergent geometry},''
  \href{http://dx.doi.org/10.1088/1126-6708/2008/02/101}{{\em JHEP} {\bf 02}
  (2008)  101},
\href{http://arxiv.org/abs/0708.0561}{{\tt arXiv:0708.0561 [hep-th]}}.

\bibitem{Nicolis:2004qq}
A.~Nicolis and R.~Rattazzi, ``{Classical and quantum consistency of the DGP
  model},'' {\em JHEP} {\bf 06} (2004)  059,
\href{http://arxiv.org/abs/hep-th/0404159}{{\tt arXiv:hep-th/0404159}}.

\bibitem{Nojiri:2005sr}
S.~Nojiri and S.~D. Odintsov, ``{Inhomogeneous equation of state of the
  universe: Phantom era, future singularity and crossing the phantom
  barrier},'' \href{http://dx.doi.org/10.1103/PhysRevD.72.023003}{{\em Phys.
  Rev.} {\bf D72} (2005)  023003},
\href{http://arxiv.org/abs/hep-th/0505215}{{\tt arXiv:hep-th/0505215}}.

\bibitem{Deffayet:2000uy}
C.~Deffayet, ``{Cosmology on a brane in Minkowski bulk},''
  \href{http://dx.doi.org/10.1016/S0370-2693(01)00160-5}{{\em Phys. Lett.} {\bf
  B502} (2001)  199--208},
\href{http://arxiv.org/abs/hep-th/0010186}{{\tt arXiv:hep-th/0010186}}.

\bibitem{Deffayet:2001pu}
C.~Deffayet, G.~R. Dvali, and G.~Gabadadze, ``{Accelerated universe from
  gravity leaking to extra dimensions},''
  \href{http://dx.doi.org/10.1103/PhysRevD.65.044023}{{\em Phys. Rev.} {\bf
  D65} (2002)  044023},
\href{http://arxiv.org/abs/astro-ph/0105068}{{\tt arXiv:astro-ph/0105068}}.

\bibitem{Vikman:2004dc}
A.~Vikman, ``{Can dark energy evolve to the phantom?},''
  \href{http://dx.doi.org/10.1103/PhysRevD.71.023515}{{\em Phys. Rev.} {\bf
  D71} (2005)  023515},
\href{http://arxiv.org/abs/astro-ph/0407107}{{\tt arXiv:astro-ph/0407107}}.

\bibitem{Dubovsky:2005xd}
S.~Dubovsky, T.~Gregoire, A.~Nicolis, and R.~Rattazzi, ``{Null energy condition
  and superluminal propagation},'' {\em JHEP} {\bf 03} (2006)  025,
\href{http://arxiv.org/abs/hep-th/0512260}{{\tt arXiv:hep-th/0512260}}.

\bibitem{Geroch:2010da}
R.~Geroch, ``{Faster Than Light?},''
\href{http://arxiv.org/abs/1005.1614}{{\tt arXiv:1005.1614 [gr-qc]}}.

\bibitem{ArmendarizPicon:2005nz}
C.~Armendariz-Picon and E.~A. Lim, ``{Haloes of k-Essence},''
  \href{http://dx.doi.org/10.1088/1475-7516/2005/08/007}{{\em JCAP} {\bf 0508}
  (2005)  007},
\href{http://arxiv.org/abs/astro-ph/0505207}{{\tt arXiv:astro-ph/0505207}}.

\bibitem{Bruneton:2006gf}
J.-P. Bruneton, ``{On causality and superluminal behavior in classical field
  theories. Applications to k-essence theories and MOND-like theories of
  gravity},'' \href{http://dx.doi.org/10.1103/PhysRevD.75.085013}{{\em Phys.
  Rev.} {\bf D75} (2007)  085013},
\href{http://arxiv.org/abs/gr-qc/0607055}{{\tt arXiv:gr-qc/0607055}}.

\bibitem{Bruneton:2007si}
J.-P. Bruneton and G.~Esposito-Farese, ``{Field-theoretical formulations of
  MOND-like gravity},''
  \href{http://dx.doi.org/10.1103/PhysRevD.76.124012}{{\em Phys. Rev.} {\bf
  D76} (2007)  124012},
\href{http://arxiv.org/abs/0705.4043}{{\tt arXiv:0705.4043 [gr-qc]}}.

\bibitem{Kang:2007vs}
J.~U. Kang, V.~Vanchurin, and S.~Winitzki, ``{Attractor scenarios and
  superluminal signals in k-essence cosmology},''
  \href{http://dx.doi.org/10.1103/PhysRevD.76.083511}{{\em Phys. Rev.} {\bf
  D76} (2007)  083511},
\href{http://arxiv.org/abs/0706.3994}{{\tt arXiv:0706.3994 [gr-qc]}}.

\bibitem{Adams:2006sv}
A.~Adams, N.~Arkani-Hamed, S.~Dubovsky, A.~Nicolis, and R.~Rattazzi,
  ``{Causality, analyticity and an IR obstruction to UV completion},'' {\em
  JHEP} {\bf 10} (2006)  014,
\href{http://arxiv.org/abs/hep-th/0602178}{{\tt arXiv:hep-th/0602178}}.

\bibitem{Mukhanov:2005bu}
V.~F. Mukhanov and A.~Vikman, ``{Enhancing the tensor-to-scalar ratio in simple
  inflation},'' {\em JCAP} {\bf 0602} (2006)  004,
\href{http://arxiv.org/abs/astro-ph/0512066}{{\tt arXiv:astro-ph/0512066}}.

\bibitem{Bessada:2009ns}
D.~Bessada, W.~H. Kinney, D.~Stojkovic, and J.~Wang, ``{Tachyacoustic
  Cosmology: An Alternative to Inflation},''
  \href{http://dx.doi.org/10.1103/PhysRevD.81.043510}{{\em Phys. Rev.} {\bf
  D81} (2010)  043510},
\href{http://arxiv.org/abs/0908.3898}{{\tt arXiv:0908.3898 [astro-ph.CO]}}.

\bibitem{Magueijo:2010zc}
J.~Magueijo, J.~Noller, and F.~Piazza, ``{Bimetric structure formation:
  non-Gaussian predictions},''
\href{http://arxiv.org/abs/1006.3216}{{\tt arXiv:1006.3216 [astro-ph.CO]}}.

\bibitem{Magueijo:2008sx}
J.~Magueijo, ``{Bimetric varying speed of light theories and primordial
  fluctuations},'' \href{http://dx.doi.org/10.1103/PhysRevD.79.043525}{{\em
  Phys. Rev.} {\bf D79} (2009)  043525},
\href{http://arxiv.org/abs/0807.1689}{{\tt arXiv:0807.1689 [gr-qc]}}.

\bibitem{Dubovsky:2006vk}
S.~L. Dubovsky and S.~M. Sibiryakov, ``{Spontaneous breaking of Lorentz
  invariance, black holes and perpetuum mobile of the 2nd kind},''
  \href{http://dx.doi.org/10.1016/j.physletb.2006.05.074}{{\em Phys. Lett.}
  {\bf B638} (2006)  509--514},
\href{http://arxiv.org/abs/hep-th/0603158}{{\tt arXiv:hep-th/0603158}}.

\bibitem{Babichev:2006vx}
E.~Babichev, V.~F. Mukhanov, and A.~Vikman, ``{Escaping from the black
  hole?},'' {\em JHEP} {\bf 09} (2006)  061,
\href{http://arxiv.org/abs/hep-th/0604075}{{\tt arXiv:hep-th/0604075}}.

\bibitem{Babichev:2007wg}
E.~Babichev, V.~Mukhanov, and A.~Vikman, ``{Looking beyond the horizon},''
\href{http://arxiv.org/abs/0704.3301}{{\tt arXiv:0704.3301 [hep-th]}}.

\bibitem{Eling:2007qd}
C.~Eling, B.~Z. Foster, T.~Jacobson, and A.~C. Wall, ``{Lorentz violation and
  perpetual motion},'' \href{http://dx.doi.org/10.1103/PhysRevD.75.101502}{{\em
  Phys. Rev.} {\bf D75} (2007)  101502},
\href{http://arxiv.org/abs/hep-th/0702124}{{\tt arXiv:hep-th/0702124}}.

\bibitem{horava}
P.~Horava, ``{Quantum Gravity at a Lifshitz Point},''
  \href{http://dx.doi.org/10.1103/PhysRevD.79.084008}{{\em Phys. Rev.} {\bf
  D79} (2009)  084008},
\href{http://arxiv.org/abs/0901.3775}{{\tt arXiv:0901.3775 [hep-th]}}.

\bibitem{bps2}
D.~Blas, O.~Pujolas, and S.~Sibiryakov, ``{Consistent Extension of Horava
  Gravity},'' \href{http://dx.doi.org/10.1103/PhysRevLett.104.181302}{{\em
  Phys. Rev. Lett.} {\bf 104} (2010)  181302},
\href{http://arxiv.org/abs/0909.3525}{{\tt arXiv:0909.3525 [hep-th]}}.

\bibitem{Blas:2010hb}
D.~Blas, O.~Pujolas, and S.~Sibiryakov, ``{Models of non-relativistic quantum
  gravity: the good, the bad and the healthy},''
\href{http://arxiv.org/abs/1007.3503}{{\tt arXiv:1007.3503 [hep-th]}}.

\bibitem{Anisimov:2005ne}
A.~Anisimov, E.~Babichev, and A.~Vikman, ``{B-inflation},''
  \href{http://dx.doi.org/10.1088/1475-7516/2005/06/006}{{\em JCAP} {\bf 0506}
  (2005)  006},
\href{http://arxiv.org/abs/astro-ph/0504560}{{\tt arXiv:astro-ph/0504560}}.

\bibitem{Caldwell:2005ai}
R.~R. Caldwell and M.~Doran, ``{Dark-energy evolution across the
  cosmological-constant boundary},''
  \href{http://dx.doi.org/10.1103/PhysRevD.72.043527}{{\em Phys. Rev.} {\bf
  D72} (2005)  043527},
\href{http://arxiv.org/abs/astro-ph/0501104}{{\tt arXiv:astro-ph/0501104}}.

\bibitem{Hu:2004kh}
W.~Hu, ``{Crossing the phantom divide: Dark energy internal degrees of
  freedom},'' \href{http://dx.doi.org/10.1103/PhysRevD.71.047301}{{\em Phys.
  Rev.} {\bf D71} (2005)  047301},
\href{http://arxiv.org/abs/astro-ph/0410680}{{\tt arXiv:astro-ph/0410680}}.

\bibitem{Zhao:2005vj}
G.-B. Zhao, J.-Q. Xia, M.~Li, B.~Feng, and X.~Zhang, ``{Perturbations of the
  Quintom Models of Dark Energy and the Effects on Observations},''
  \href{http://dx.doi.org/10.1103/PhysRevD.72.123515}{{\em Phys. Rev.} {\bf
  D72} (2005)  123515},
\href{http://arxiv.org/abs/astro-ph/0507482}{{\tt arXiv:astro-ph/0507482}}.

\bibitem{Sen:2005ra}
A.~A. Sen, ``{Reconstructing K-essence},'' {\em JCAP} {\bf 0603} (2006)  010,
\href{http://arxiv.org/abs/astro-ph/0512406}{{\tt arXiv:astro-ph/0512406}}.

\bibitem{Abramo:2005be}
L.~R. Abramo and N.~Pinto-Neto, ``{On the Stability of Phantom K-essence
  Theories},'' \href{http://dx.doi.org/10.1103/PhysRevD.73.063522}{{\em Phys.
  Rev.} {\bf D73} (2006)  063522},
\href{http://arxiv.org/abs/astro-ph/0511562}{{\tt arXiv:astro-ph/0511562}}.

\bibitem{Kunz:2006wc}
M.~Kunz and D.~Sapone, ``{Crossing the Phantom Divide},''
  \href{http://dx.doi.org/10.1103/PhysRevD.74.123503}{{\em Phys. Rev.} {\bf
  D74} (2006)  123503},
\href{http://arxiv.org/abs/astro-ph/0609040}{{\tt arXiv:astro-ph/0609040}}.

\bibitem{Feng:2004ad}
B.~Feng, X.-L. Wang, and X.-M. Zhang, ``{Dark Energy Constraints from the
  Cosmic Age and Supernova},''
  \href{http://dx.doi.org/10.1016/j.physletb.2004.12.071}{{\em Phys. Lett.}
  {\bf B607} (2005)  35--41},
\href{http://arxiv.org/abs/astro-ph/0404224}{{\tt arXiv:astro-ph/0404224}}.

\bibitem{Cai:2009zp}
Y.-F. Cai, E.~N. Saridakis, M.~R. Setare, and J.-Q. Xia, ``{Quintom Cosmology:
  Theoretical implications and observations},''
  \href{http://dx.doi.org/10.1016/j.physrep.2010.04.001}{{\em Phys. Rept.} {\bf
  493} (2010)  1--60},
\href{http://arxiv.org/abs/0909.2776}{{\tt arXiv:0909.2776 [hep-th]}}.

\bibitem{Zhang:2009dw}
H.~Zhang, ``{Crossing the phantom divide},''
\href{http://arxiv.org/abs/0909.3013}{{\tt arXiv:0909.3013 [astro-ph.CO]}}.

\bibitem{Copeland:2006wr}
E.~J. Copeland, M.~Sami, and S.~Tsujikawa, ``{Dynamics of dark energy},''
  \href{http://dx.doi.org/10.1142/S021827180600942X}{{\em Int. J. Mod. Phys.}
  {\bf D15} (2006)  1753--1936},
\href{http://arxiv.org/abs/hep-th/0603057}{{\tt arXiv:hep-th/0603057}}.

\bibitem{Boisseau:2000pr}
B.~Boisseau, G.~Esposito-Farese, D.~Polarski, and A.~A. Starobinsky,
  ``{Reconstruction of a scalar-tensor theory of gravity in an accelerating
  universe},'' \href{http://dx.doi.org/10.1103/PhysRevLett.85.2236}{{\em Phys.
  Rev. Lett.} {\bf 85} (2000)  2236},
\href{http://arxiv.org/abs/gr-qc/0001066}{{\tt arXiv:gr-qc/0001066}}.

\bibitem{Gannouji:2006jm}
R.~Gannouji, D.~Polarski, A.~Ranquet, and A.~A. Starobinsky, ``{Scalar-tensor
  models of normal and phantom dark energy},'' {\em JCAP} {\bf 0609} (2006)
  016,
\href{http://arxiv.org/abs/astro-ph/0606287}{{\tt arXiv:astro-ph/0606287}}.

\bibitem{Hu:2007nk}
W.~Hu and I.~Sawicki, ``{Models of f(R) Cosmic Acceleration that Evade
  Solar-System Tests},''
  \href{http://dx.doi.org/10.1103/PhysRevD.76.064004}{{\em Phys. Rev.} {\bf
  D76} (2007)  064004},
\href{http://arxiv.org/abs/0705.1158}{{\tt arXiv:0705.1158 [astro-ph]}}.

\bibitem{Motohashi:2010tb}
H.~Motohashi, A.~A. Starobinsky, and J.~Yokoyama, ``{Phantom boundary crossing
  and anomalous growth index of fluctuations in viable f(R) models of cosmic
  acceleration},''
\href{http://arxiv.org/abs/1002.1141}{{\tt arXiv:1002.1141 [astro-ph.CO]}}.

\bibitem{Lim:2010yk}
E.~A. Lim, I.~Sawicki, and A.~Vikman, ``{Dust of Dark Energy},''
  \href{http://dx.doi.org/10.1088/1475-7516/2010/05/012}{{\em JCAP} {\bf 1005}
  (2010)  012},
\href{http://arxiv.org/abs/1003.5751}{{\tt arXiv:1003.5751 [astro-ph.CO]}}.

\bibitem{Creminelli:2008wc}
P.~Creminelli, G.~D'Amico, J.~Norena, and F.~Vernizzi, ``{The Effective Theory
  of Quintessence: the $w<-1$ Side Unveiled},''
  \href{http://dx.doi.org/10.1088/1475-7516/2009/02/018}{{\em JCAP} {\bf 0902}
  (2009)  018},
\href{http://arxiv.org/abs/0811.0827}{{\tt arXiv:0811.0827 [astro-ph]}}.

\bibitem{Freese:2002sq}
K.~Freese and M.~Lewis, ``{Cardassian Expansion: a Model in which the Universe
  is Flat, Matter Dominated, and Accelerating},''
  \href{http://dx.doi.org/10.1016/S0370-2693(02)02122-6}{{\em Phys. Lett.} {\bf
  B540} (2002)  1--8},
\href{http://arxiv.org/abs/astro-ph/0201229}{{\tt arXiv:astro-ph/0201229}}.

\bibitem{Koivisto:2005mm}
T.~Koivisto and D.~F. Mota, ``{Dark Energy Anisotropic Stress and Large Scale
  Structure Formation},''
  \href{http://dx.doi.org/10.1103/PhysRevD.73.083502}{{\em Phys. Rev.} {\bf
  D73} (2006)  083502},
\href{http://arxiv.org/abs/astro-ph/0512135}{{\tt arXiv:astro-ph/0512135}}.

\bibitem{Mota:2007sz}
D.~F. Mota, J.~R. Kristiansen, T.~Koivisto, and N.~E. Groeneboom,
  ``{Constraining Dark Energy Anisotropic Stress},''
  \href{http://dx.doi.org/10.1111/j.1365-2966.2007.12413.x}{{\em Mon. Not. Roy.
  Astron. Soc.} {\bf 382} (2007)  793--800},
\href{http://arxiv.org/abs/0708.0830}{{\tt arXiv:0708.0830 [astro-ph]}}.

\bibitem{Koivisto:2008ig}
T.~Koivisto and D.~F. Mota, ``{Anisotropic Dark Energy: Dynamics of Background
  and Perturbations},''
  \href{http://dx.doi.org/10.1088/1475-7516/2008/06/018}{{\em JCAP} {\bf 0806}
  (2008)  018},
\href{http://arxiv.org/abs/0801.3676}{{\tt arXiv:0801.3676 [astro-ph]}}.

\bibitem{dePutter:2010vy}
R.~de~Putter, D.~Huterer, and E.~V. Linder, ``{Measuring the Speed of Dark:
  Detecting Dark Energy Perturbations},''
  \href{http://dx.doi.org/10.1103/PhysRevD.81.103513}{{\em Phys. Rev.} {\bf
  D81} (2010)  103513},
\href{http://arxiv.org/abs/1002.1311}{{\tt arXiv:1002.1311 [astro-ph.CO]}}.

\bibitem{Doran:2001rw}
M.~Doran, J.-M. Schwindt, and C.~Wetterich, ``{Structure formation and the time
  dependence of quintessence},''
  \href{http://dx.doi.org/10.1103/PhysRevD.64.123520}{{\em Phys. Rev.} {\bf
  D64} (2001)  123520},
\href{http://arxiv.org/abs/astro-ph/0107525}{{\tt arXiv:astro-ph/0107525}}.

\bibitem{Doran:2007ep}
M.~Doran, G.~Robbers, and C.~Wetterich, ``{Impact of three years of data from
  the Wilkinson Microwave Anisotropy Probe on cosmological models with
  dynamical dark energy},''
  \href{http://dx.doi.org/10.1103/PhysRevD.75.023003}{{\em Phys. Rev.} {\bf
  D75} (2007)  023003},
\href{http://arxiv.org/abs/astro-ph/0609814}{{\tt arXiv:astro-ph/0609814}}.

\bibitem{Xia:2009ys}
J.-Q. Xia and M.~Viel, ``{Early Dark Energy at High Redshifts: Status and
  Perspectives},'' \href{http://dx.doi.org/10.1088/1475-7516/2009/04/002}{{\em
  JCAP} {\bf 0904} (2009)  002},
\href{http://arxiv.org/abs/0901.0605}{{\tt arXiv:0901.0605 [astro-ph.CO]}}.

\bibitem{dePutter:2009kn}
R.~de~Putter, O.~Zahn, and E.~V. Linder, ``{CMB Lensing Constraints on
  Neutrinos and Dark Energy},''
  \href{http://dx.doi.org/10.1103/PhysRevD.79.065033}{{\em Phys. Rev.} {\bf
  D79} (2009)  065033},
\href{http://arxiv.org/abs/0901.0916}{{\tt arXiv:0901.0916 [astro-ph.CO]}}.

\bibitem{Hollenstein:2009ph}
L.~Hollenstein, D.~Sapone, R.~Crittenden, and B.~M. Schaefer, ``{Constraints on
  early dark energy from CMB lensing and weak lensing tomography},''
  \href{http://dx.doi.org/10.1088/1475-7516/2009/04/012}{{\em JCAP} {\bf 0904}
  (2009)  012},
\href{http://arxiv.org/abs/0902.1494}{{\tt arXiv:0902.1494 [astro-ph.CO]}}.

\bibitem{Wang:2008zh}
Y.~Wang, ``{Figure of Merit for Dark Energy Constraints from Current
  Observational Data},''
  \href{http://dx.doi.org/10.1103/PhysRevD.77.123525}{{\em Phys. Rev.} {\bf
  D77} (2008)  123525},
\href{http://arxiv.org/abs/0803.4295}{{\tt arXiv:0803.4295 [astro-ph]}}.

\bibitem{Kobayashi:2010cm}
T.~Kobayashi, M.~Yamaguchi, and J.~Yokoyama, ``{G-inflation: inflation driven
  by the Galileon field},''
\href{http://arxiv.org/abs/1008.0603}{{\tt arXiv:1008.0603 [hep-th]}}.

\bibitem{Maldacena:2002vr}
J.~M. Maldacena, ``{Non-Gaussian features of primordial fluctuations in single
  field inflationary models},'' {\em JHEP} {\bf 05} (2003)  013,
\href{http://arxiv.org/abs/astro-ph/0210603}{{\tt arXiv:astro-ph/0210603}}.

\end{thebibliography}\endgroup
\end{document}